\documentclass[aps,prl,longbibliography,twocolumn,superscriptaddress]{revtex4-2}
\usepackage{amsmath,amssymb,bm,graphicx,color,gensymb,bbold,hyperref,keyval,url,latexsym}
\usepackage{xcolor}
\usepackage[normalem]{ulem} 
\usepackage{CJK}
\usepackage{float}

\begin{document}
\title{Where is the Quantum Spin Nematic?}

\begin{CJK*}{UTF8}{}
\author{Shengtao Jiang (\CJKfamily{gbsn}蒋晟韬)}
\affiliation{Department of Physics and Astronomy, University of California, Irvine, California 92697, USA}
\author{Judit Romh\'anyi}
\affiliation{Department of Physics and Astronomy, University of California, Irvine, California 92697, USA}
\author{Steven R. White}
\affiliation{Department of Physics and Astronomy, University of California, Irvine, California 92697, USA}
\author{M. E. Zhitomirsky}
\affiliation{Universit\'e Grenoble Alpes, Grenoble INP, CEA, IRIG, PHELIQS, 38000 Grenoble, France}
\affiliation{Institut Laue-Langevin, 71 Avenue des Martyrs, CS 20156, 38042 Grenoble Cedex 9, France}
\author{A. L. Chernyshev}
\affiliation{Department of Physics and Astronomy, University of California, Irvine, California 92697, USA}

\date{\today}

\begin{abstract}
We provide strong evidence of the spin-nematic state in a paradigmatic ferro-antiferromagnetic  $J_1$--$J_2$ model using analytical and density-matrix renormalization group methods. In zero field, the attraction of spin-flip pairs leads to a first-order transition and no nematic state, while  pair-repulsion at larger $J_2$ stabilizes the nematic phase in a narrow region near the pair-condensation field. A devil's staircase of  multi-pair condensates is conjectured for   weak pair-attraction. A suppression of the spin-flip gap by  many-body effects leads to an order-of-magnitude contraction of the nematic phase compared to  na\"ive expectations. The proposed phase diagram should be broadly valid.
\end{abstract}
\maketitle
\end{CJK*}

{\it Introduction.}---Liquid crystals---which combine properties of a liquid and a solid that seem mutually exclusive---were considered an exotic state of matter for nearly a century before becoming ubiquitous in technology~\cite{deGennes1993,Andrienko_18}. Their quantum analogues have been hypothesized and pursued in several contexts, such as electronic nematic states in strongly correlated materials~\cite{cuprate2,iron2,iron1,cuprate1}, spin nematics in frustrated magnets~\cite{Mila_17,exp2008sq,exp2011licuvo4,exp2017sq,exp2017licuvo4,exp2019sq,zheludev1,zheludev2,
zheludev3,zheludev4,zheludev5}, and supersolids in He$^4$ and cold atomic gases~\cite{Andreev_69,Leggett_70,supersolid1,supersolid2}.
Quantum spin nematics are particularly elusive, as they should interpolate between a magnetically ordered spin solid and a spin liquid, another exotic and elusive state~\cite{Savary_2016,Knolle_19}.
Like spin liquids, spin nematics lack  conventional  dipolar magnetic order, but instead break spin-rotational symmetry with  quadrupolar or higher-rank multipolar ordering \cite{1969nm,1984nm,chubukov1991}, making their experimental detection challenging \cite{PencLauchli2011}.

An earlier study   has proposed an intuitive view of the nematic states as of the Bose-Einstein condensates (BECs) of {\it pairs} of spin excitations with a gap in the single-particle sector \cite{chubukov1991}. In a nutshell, a nematic state  occurs if a conventional order due to a BEC of single spin flips \cite{Batista_14} is preempted by a BEC of their pairs. Since the bound states (BSs) of magnons in  ferromagnets (FMs)  do not Bose-condense \cite{Wortis_63,Mattis}, it was suggested that magnetic frustration can facilitate nematic pair-BEC \cite{chubukov1991}, a concept  explored   in several classes of  frustrated magnets  theoretically 
\cite{shannon2006nematic,Sindzingre_2009,Igbal_16,kuzian2007,momoi1,momoi2,momoi3,momoi4,Lauchli09,%
oleg1,zhitomirsky2010,ueda2009magnon,Janson2016,MilaSS_05,UhrigSS_00,UedaSS_01} and experimentally \cite{exp2011licuvo4,exp2017licuvo4,exp2008sq,exp2017sq,exp2019sq,zheludev1,zheludev2,zheludev3,zheludev4,%
zheludev5}.

\begin{figure}[t]
\centering    
\includegraphics[width=1.0\columnwidth]{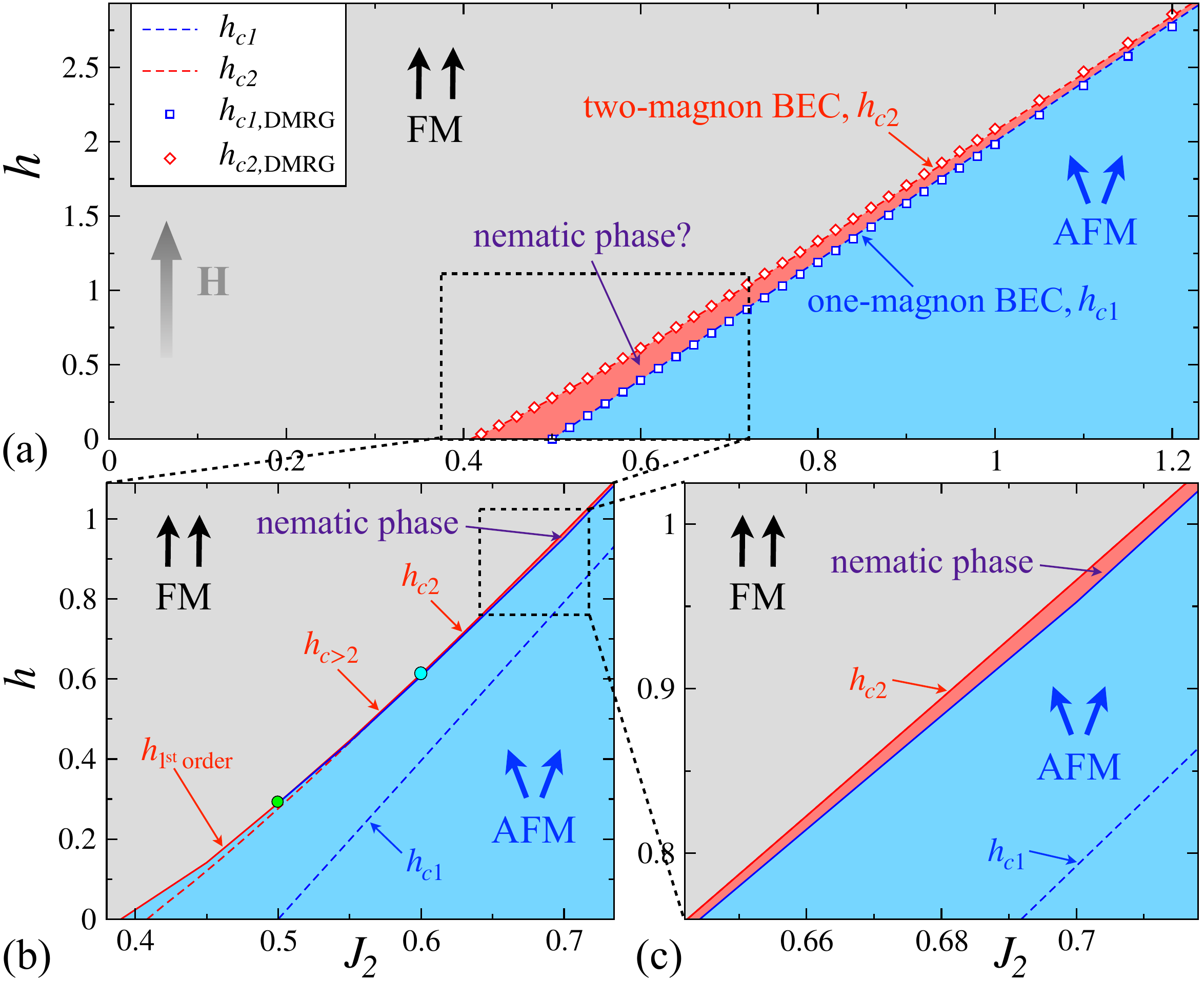}
\vskip -0.25cm
\caption{(a) The na\"{i}ve $h$--$J_2$ phase diagram of  model (\ref{eq:j1j2mod}) based on the single spin-flip and pair-BEC $h_{c1}$  and $h_{c2}$ lines.  Lines and symbols show analytical and DMRG results, respectively.
(b) The actual phase diagram of  the model (\ref{eq:j1j2mod}) in the zoomed region of (a), with the first-order, multi-pair, and pair-BEC transitions emphasized. (c) The zoomed sector of (b) showing the  extent of the nematic phase  near  pair-BEC field.}
\label{fig:pairing}
\vskip -0.62cm
\end{figure}

One of the simplest paradigmatic models for this scenario is  the   $J_1$--$J_2$ ferro-antiferromagnetic (AFM) $S\!=\!1/2$ Heisenberg model on a square lattice in external field, 
\vskip -0.15cm
\noindent
\begin{equation}
    \label{eq:j1j2mod}
    H=J_1\sum_{\langle ij \rangle_1}~{{\bf S}_i}\cdot {{\bf S}_j} 
    +J_2\sum_{\langle ij \rangle_2} 
    {{\bf S}_i}\cdot {{\bf S}_j} -h\sum_i S^z_i,
\end{equation}
\vskip -0.25cm
\noindent
where $\langle ij \rangle_{1(2)}$ denotes the first (second) nearest-neighbor bonds, the field $h\!=\!g\mu_BH$,  $J_1\!=\!-1$ is set as the energy unit, and  $J_2\!>\!0$. The FM is a ground state for small $J_2$; for large $J_2$ it is a stripe AFM \cite{edzerofield}; see Fig.~\ref{fig:pairing}(a).

Prior studies on this model \cite{Sindzingre_2009,shannon2006nematic,Igbal_16} have proposed  the nematic state to intervene between FM and AFM phases in a broad region similar to the one shown in Fig.~\ref{fig:pairing}(a). However, this contradicts the robust numerical evidence of a direct FM-AFM transition in zero field \cite{edzerofield}, highlighting a common pitfall of claiming the nematic state based on correlations that are subsidiary to a prevalent dipolar order. It also shows that the nematic state of BEC pairs may be  superseded by other instabilities.

In this Letter, we combine analytical and numerical density-matrix renormalization group (DMRG) approaches to provide unambiguous conclusions on the nematic state in the $J_1$--$J_2$ square-lattice model. 

\begin{figure*}[t]
\centering    
\includegraphics[width=2.0\columnwidth]{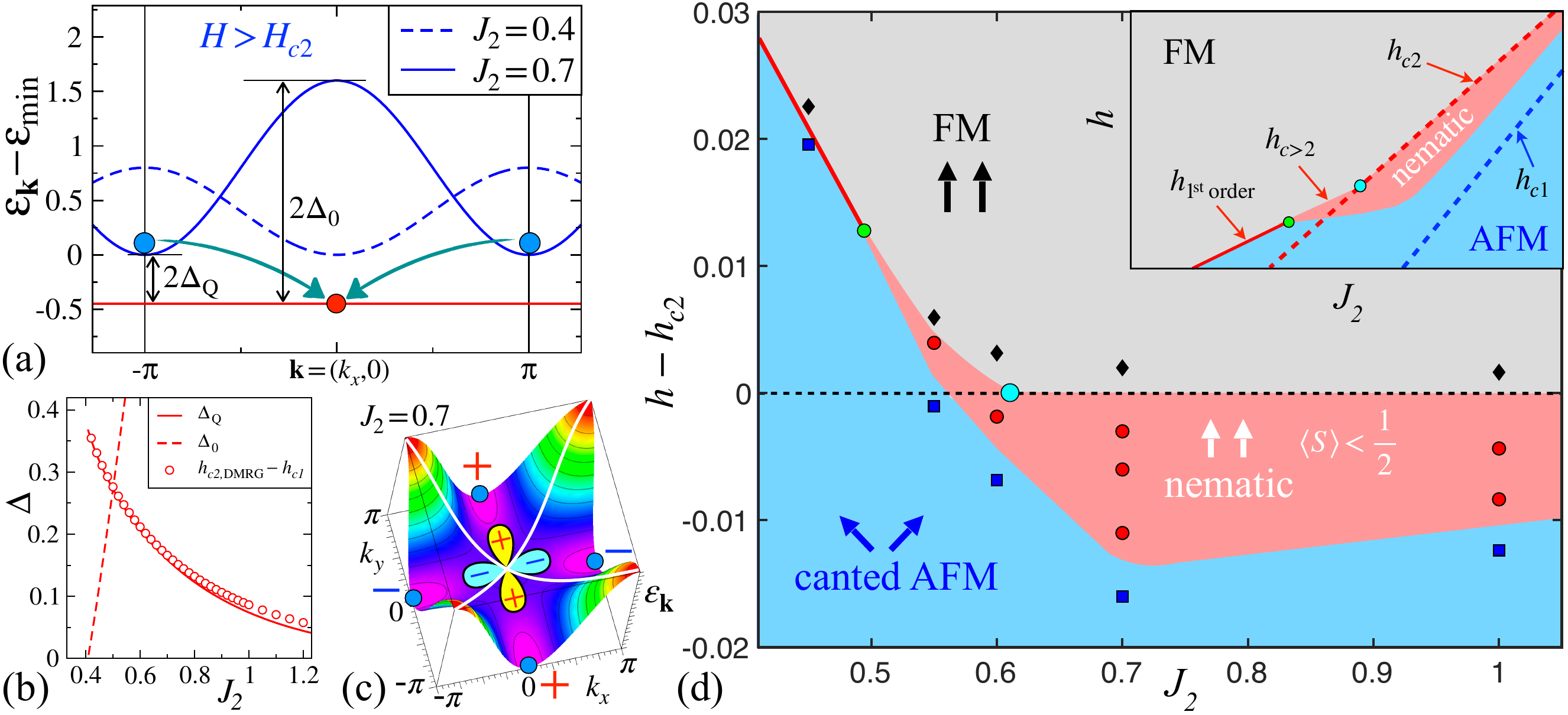}
\vskip -0.4cm
\caption{(a) Magnon energies $\varepsilon_{\bf k}$ at $h\!>\!h_{c2}$ for $J_2\!=\!0.7$ and 0.4, schematics of magnon pairing, and  gaps $\Delta_{{\bf Q}(0)}$. (b) The pairing gap $\Delta$ vs $J_2$ from  theory (lines)   and DMRG (symbols). (c)  $\varepsilon_{\bf k}$ for $J_2\!=\!0.7$, nodes of the $d_{x^2-y^2}$-wave harmonic (white lines), and schematics of the $d$-wave. (d) The $h$--$J_2$ phase diagram of the model (\ref{eq:j1j2mod}) by DMRG, field $h$ is relative to $h_{c2}$.  Symbols mark the FM (black), nematic (red), and AFM (blue) phases.  Phase boundaries are inferred from the midpoints between the data. Cyan circle marks a switch to the pair-attraction and green circle to the first-order transition (solid line). Inset: Schematics of the true $h$--$J_2$ phase diagram in Fig.~\ref{fig:pairing}(b). The nematic region and the deviation from the $h_{c2}$-line are exaggerated.}
\label{fig:phasediag}
\vskip -0.6cm
\end{figure*}

{\it D-wave pair-BEC.}---Pairing is ubiquitous  in physics \cite{Annett,Taylor}. In  model (\ref{eq:j1j2mod}), the pairing of  two spin flips sharing an attractive FM $J_1$-link occurs  in the polarized state. Since the model is 2D, one expects a BS in the $s$-wave channel for an arbitrarily weak attraction, or any $J_2$, as in the Cooper problem for superconductivity \cite{Annett}. Yet, the prior works give a finite $J_2$-range for the pairing \cite{Sindzingre_2009,shannon2006nematic} and provide no insight into the  pairs' $d$-wave symmetry. 

The paring of two spin flips can be solved by an exact formalism \cite{Wortis_63,zhitomirsky2010}. It yields the na\"{i}ve phase diagram of the model (\ref{eq:j1j2mod}) shown in Fig.~\ref{fig:pairing}(a), where $h_{c1}\!=\!4J_2-2$ is the line of the single spin-flip BEC and the FM-AFM border in the classical limit, which is preempted by the pair-BEC at $h_{c2}$ for {\it any} $J_2$. DMRG energies for $16\times 8$ cylinders with fixed numbers of spin flips yield $h_{c1}$ and $h_{c2}$ values in nearly-perfect agreement (symbols).

The magnon pairing gap $\Delta$, sketched in Fig.~\ref{fig:phasediag}(a), is the difference of these fields, $\Delta\!\equiv\!h_{c2}\!-\!h_{c1}$,  which agrees with the weak-coupling result of the Cooper problem \cite{Annett}
\vskip -0.17cm
\noindent
\begin{equation}
    \label{eq:Delta}
   \Delta \approx J_2 \,e^{-\pi J_2},
\end{equation}
\vskip -0.17cm
\noindent
for $J_2\!\gg\!1$, but in the $d$-wave channel. Fig.~\ref{fig:phasediag}(c) explains the predominance of the $d$-wave. The nodes of the $d_{x^2-y^2}$ harmonic of the attraction potential, $V_{\bf q}^d\!\propto\!(\cos q_x\!-\!\cos q_y)$, avoid crossing the magnon band minima at 
${\bf Q}\!=\!(0,\pi) [(\pi,0)]$, 
see Fig.~\ref{fig:phasediag}(a), while the nodes of  other harmonics  do cross them, rendering pairing in these channels unfavorable  \cite{sm}.
The spatial extent of the BS in (\ref{eq:Delta}) can be estimated 
as $\xi\!\propto\!\sqrt{J_2/\Delta}\!\propto\!e^{\pi J_2/2}$, relating deviations of the 
DMRG from exact results in Fig.~\ref{fig:phasediag}(b) at larger $J_2$ to the finite-size effect
\footnote{For example,  $\xi\!\approx\!4.8$ for $J_2\!=\!1.0$ \cite{sm}.}.

{\it Phase diagram.}---With the pairing problem in the FM state solved exactly,  its $d$-wave symmetry and $J_2$-extent elucidated, a nematic phase is expected to exist below the pair-BEC transition $h_{c2}$ down to the single spin-flip BEC  $h_{c1}$,  where the single-particle gap closes and the AFM order prevails, see the phase diagram in Fig.~\ref{fig:pairing}(a). However, as we demonstrate, the many-body effects strongly alter some of these expectations, see Figs.~\ref{fig:pairing}(b), \ref{fig:pairing}(c), and \ref{fig:phasediag}(d). 

Generally, for a BEC condensate to form a 
superfluid phase its constituents must repel  \cite{Lifshitz_v2,Batista_14}. This is 
the case for the pair-BEC 
for large (repulsive) $J_2$,
implying 
that the nematic phase {\it must} occur in {\it some} region below the $h_{c2}$-line, which  
is unaffected by  many-body effects. 

As the pair binding energy $2\Delta$ increases for smaller $J_2$, see Fig.~\ref{fig:phasediag}(b), one also expects a change of the {\it pair-pair} interaction from repulsive to attractive. With the numerical evidence for that presented below, this change occurs at about $J_2\!\approx\!0.6$, marked by a cyan circle in the  phase diagrams in Figs.~\ref{fig:pairing}(b) and \ref{fig:phasediag}(d). 

The pair-attraction has two effects.  First, the FM-nematic phase boundary in Figs.~\ref{fig:phasediag}(d) and \ref{fig:pairing}(b) is pulled above the $h_{c2}$-line, superseded by  a BEC of the multi-pair states~\footnote{The ``true'' multi-pair states are likely to occupy only a very narrow region below the $h_{c>2}$ phase boundary, while the nematic phase is also pulled above $h_{c2}$, in analogy to the AFM state that expands above the $h_{c1}$-line.}. Second, the nematic region shrinks  as the critical pair density  for a transition to the dipolar state is reached more readily. Ultimately, at about $J_2\!\approx\!0.5$ (green circle in Figs.~\ref{fig:phasediag}(d) and \ref{fig:pairing}(b)), the nematic phase ceases altogether. In a sense, while the pair-binding gets stronger, the stiffness of the  phase vanishes, leading to a first-order collapse of the  FM  into AFM phase with a finite canting of spins, explaining the zero-field results of Ref.~\cite{edzerofield} and substantiating the proposal of Ref.~\cite{ueda_phasesep}.

The most striking change concerns the na\"{i}ve nematic-AFM phase boundary in Fig.~\ref{fig:pairing}(a). The $h_{c1}$-line corresponds to a closing of the single-magnon  gap for the  non-interacting magnons.  However, in the presence of the pair-BEC, this gap is strongly reduced due to attraction to the pair condensate~\cite{sm},  dramatically extending the AFM phase {\it above} the $h_{c1}$-line and leading  to about an order-of-magnitude contraction of the na\"{i}ve nematic phase according to DMRG \cite{footnote}; see Figs.~\ref{fig:pairing} and \ref{fig:phasediag}(d).

Our Fig.~\ref{fig:phasediag}(d)  and Figs.~\ref{fig:pairing}(b) and \ref{fig:pairing}(c) quantify all of the trends described above: the  narrow  nematic region below the $h_{c2}$-line, the change to the pair-attractive regime for  $J_2\!\alt\!0.6$ leading to multi-pair transitions and further narrowing of the nematic region, and  first-order transition for $J_2\!\alt\!0.5$ together with a shift of the FM-to-AFM boundary from the $h_{c2}$-line to smaller $J_2$.

To reveal the resultant phase diagram in Figs.~\ref{fig:pairing}(b) and \ref{fig:pairing}(c), we use iterative zooming because the width of the nematic region and the shift of the transition lines are hard to discern on the scale of Fig.~\ref{fig:pairing}(a). They are derived from Figure~\ref{fig:phasediag}(d), which is based on the DMRG results discussed below, with each symbol corresponding to an individual simulation. 

{\it DMRG results.}---DMRG calculations are performed on the $L_x\!\times\!L_y$-site square-lattice cylinders with mixed boundary conditions, 
and  width $L_y\!=\!8$. 
\footnote{Calculations are carried out using the ITensor  library \cite{itensor}, typically performing 16 sweeps and reaching a maximum bond dimension of about $m\!=\!2000$ to ensure good convergence with a truncation error  $<\!10^{-6}$. To avoid metastable states, we use different initial spin configurations and compare converged energies to ensure the ground state is reached.}

We use three complementary approaches. The first is long-cylinder ``scans,'' in which the magnetic field is varied along the length of the $40\!\times\!8$ cylinder,  with different phases and their boundaries coexisting in one system. These 1D cuts through the phase diagram are very useful \cite{scan1,scan2,scan3,scan4,scan5}, 
allowing one to differentiate first- and second-order transitions by varying the ranges of the scans. Since the parameter gradient can impose unwanted proximity effects, we use such scans judiciously as the first exploratory measure  of the nematic phase.

The second approach utilizes  $20\!\times\!8$ cylinders, with an aspect ratio that 
approximates
the 2D behavior in the thermodynamic limit~\cite{FS}. To obtain BEC boundaries in Fig.~\ref{fig:pairing}, the pairing gap in  Fig.~\ref{fig:phasediag}(b), and multi-pair energies, we perform calculations for fixed numbers of spin flips (fixed total $S^z$) as a function of $h$ and $J_2$. 

Lastly, the  same cylinders are simulated without fixing total $S^z$ to allow for symmetry-broken phases that are induced by weak edge fields. The broken symmetry  allows us to measure local order parameters 
 instead of their 
 correlation functions \cite{scan1,scan2,scan3,scan4,scan5}. The decay of the induced orders away from the boundary also serves as an excellent indicator of their stability in the 2D bulk.

\begin{figure}[t]
\centering    
\includegraphics[width=1.0\columnwidth]{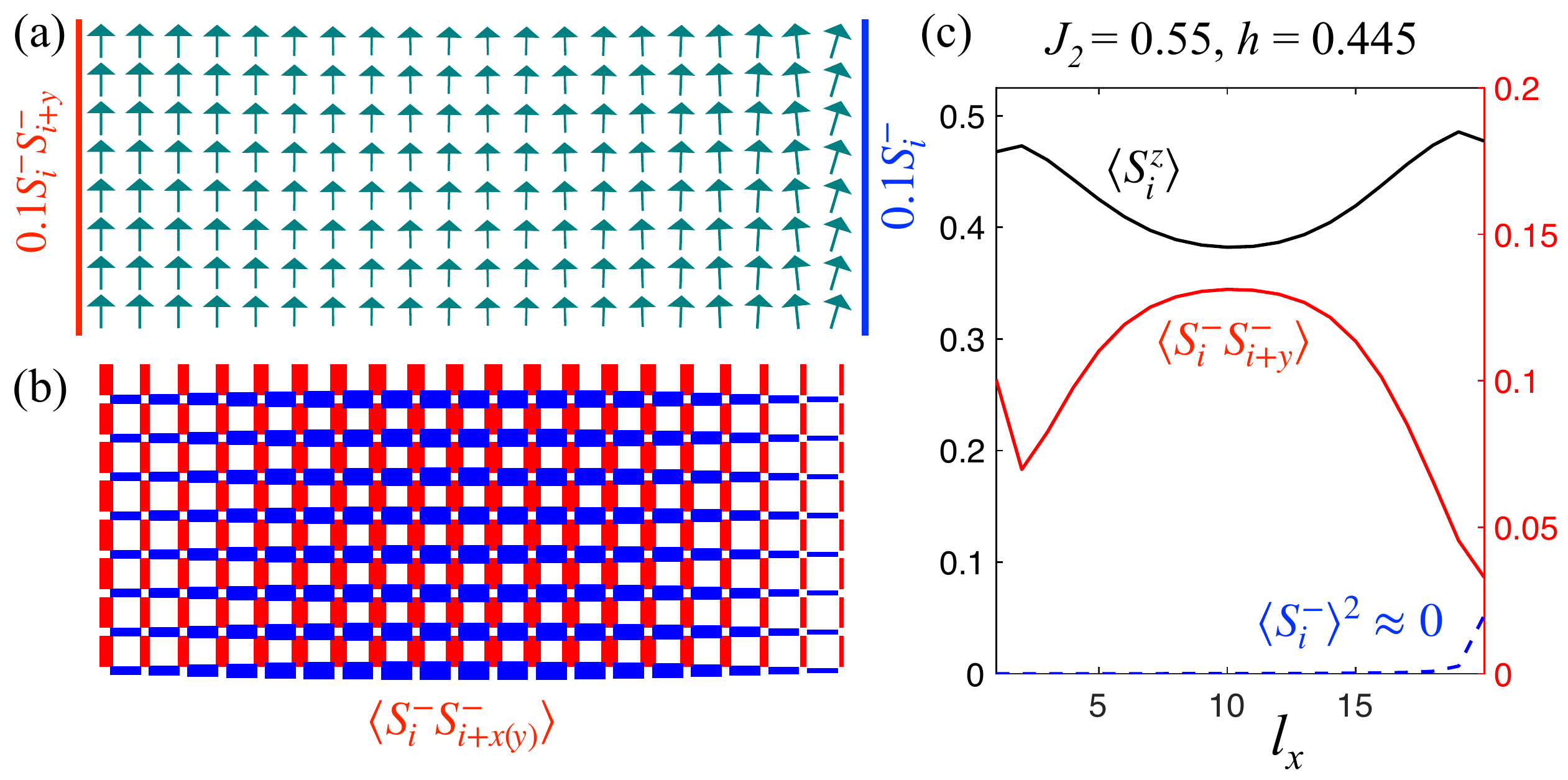}
\vskip -0.35cm
\caption{DMRG results in the $20\!\times\!8$ cluster for $J_2\!=\!0.55$ and $h\!=\!0.445$. (a) Ordered moment $\langle S \rangle$ in the $xz$-plane with pairing field $0.1 S^-_i S^-_{i+y}$ (spin-flip field $0.1S^-_i$) at the left (right) edge. (b) Nearest-neighbor component of the pair wave-function; thickness (color) of the bond corresponds to the value (sign) of $\langle S^-_i S^-_{i+{x(y)}}\rangle$. (c) $z$-axis magnetization $\langle S^z_i \rangle\!\approx\!\langle S \rangle$ (left axis), and nematic $\langle S^-_i S^-_{i+y}\rangle$ and spin-canting $\langle S^-_i \rangle^2$ order parameters (right axis) along the cylinder.}
\label{fig:nematic}
\vskip -0.6cm
\end{figure}

Our Figure~\ref{fig:nematic} showcases the described approach and its results for $J_2\!=\!0.55$ and $h\!=\!0.445$; see the leftmost red circle in Fig.~\ref{fig:phasediag}(d), just above $h_{c2}\!=\!0.441$ for this value of $J_2$.  Fig.~\ref{fig:nematic}(a) shows the spin configuration, with arrows' length equal to the local ordered moment $\langle S \rangle$. In Fig.~\ref{fig:nematic}(b)  bonds represent the nearest-neighbor pair wave-function $\langle S^-_i S^-_{i+{x(y)}}\rangle$, which is directly related to the quadrupole-moment order parameter \cite{Lauchli09}, and  Fig.~\ref{fig:nematic}(c) provides a quantitative measure of them along the length of the cluster. A pairing field $0.1 S^-_i S^-_{i+y}$ (spin-flip field $0.1S^-_i$) is applied at the left (right) edge.
  
In order to avoid the pitfalls of the earlier work \cite{shannon2006nematic}, an important step in the search for the nematics is  to rigorously rule out dipolar orders, since  nematic correlations also exist in them as a subsidiary of the multipole expansion. As one can see in Fig.~\ref{fig:nematic}(a) and \ref{fig:nematic}(c),  the  magnetization is markedly suppressed from  full saturation  away from the boundary, $\langle S^z \rangle \!<\!\frac12$,  but shows no sign of canting. In the same region, the quadrupolar order parameter is clearly developed, with $\langle S^-_i S^-_{i+{y}}\rangle \!\agt\! 0.1$ and its  $d$-wave character  evident from the opposite sign of the horizontal and vertical bonds in Fig.~\ref{fig:nematic}(b). On the other hand, the induced canting on the right edge decays away from it with no detectable $\langle S^-_i \rangle$ in the bulk; see Figs.~\ref{fig:nematic}(a) and \ref{fig:nematic}(c), which indicate a gap to one-magnon excitations and the absence of the dipolar order.  

Altogether, the analysis presented in Fig.~\ref{fig:nematic} leaves no doubt for the presence of the $d$-wave nematic state for the chosen values of $h$ and $J_2$. We point out again that without the pinning field, the nematic state still exists and can  be detected through the pair-pair correlations instead of the local order parameter, but they are no more informative and less visual than the results in Fig.~\ref{fig:nematic}.   

\begin{figure}[t]
\centering    
\includegraphics[width=1.0\columnwidth]{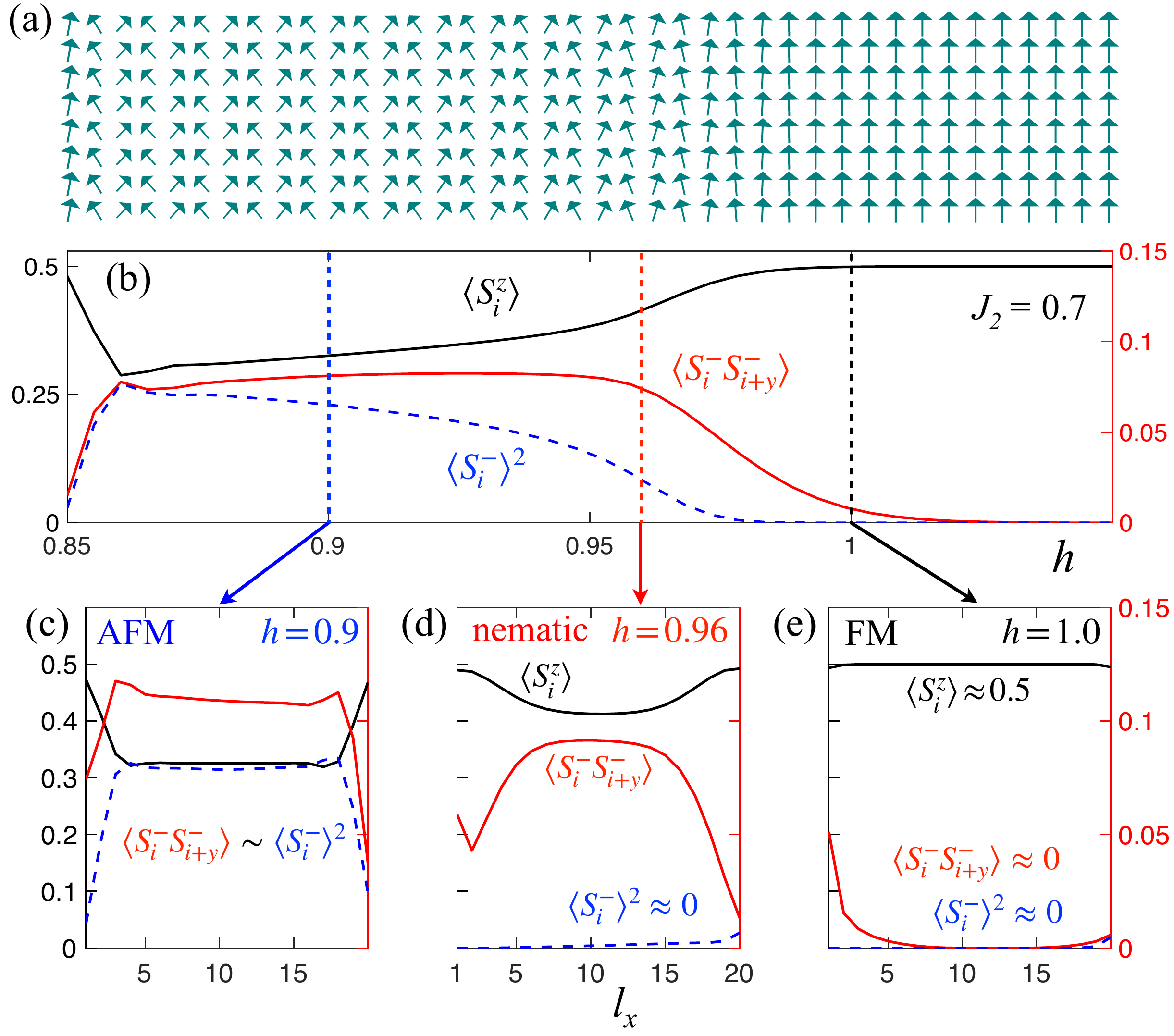}
\vskip -0.35cm
\caption{Long-cylinder scan in $h$ from 0.85 to 1.05 for $J_2\!=\!0.7$, with (a) spin pattern of the ordered moments (field $0.1S^-_i$ at the left edge), and (b) magnetization $\langle S^z_i \rangle$ (left axis), and pair $\langle S^-_i S^-_{i+y}\rangle$ and spin-canting $\langle S^-_i \rangle^2$ order parameters (right axis). (c), (d) and (e) Fixed-parameter calculations as in Fig.~\ref{fig:nematic}(c) for  $h\!=\!0.9$, 0.96, and 1.0, respectively.}
\label{fig:j20.7scan}
\vskip -0.6cm
\end{figure}

In Figure~\ref{fig:j20.7scan}, we show a long-cylinder scan for $J_2\!=\!0.7$ with varied $h$. From  Fig.~\ref{fig:pairing}(a) one  expects to see the nematic phase from the single-magnon-BEC  to the pair-BEC fields, from $h_{c1}\!=\!0.792$ to $h_{c2}\!=\!0.966$.  Instead, we observe a robust  AFM phase with substantial dipolar order $\langle S^-_i \rangle$ all the way up to a  vicinity of $h_{c2}$; see Figs.~\ref{fig:j20.7scan}(a) and \ref{fig:j20.7scan}(b). Although $\langle S^z \rangle$   in Fig.~\ref{fig:j20.7scan}(b) drops precipitously in a narrow field range near $h_{c2}$, varying the limits of the scan suggests second-order transition(s).

Fig.~\ref{fig:j20.7scan}(b) shows that near $h_{c2}$  the nematic order parameter dominates the dipolar one, suggesting the presence of the nematic phase. This behavior is markedly different from the case of the quadrupolar order  occurring  as a byproduct of the dipolar one in the pure AFM model~\cite{sm}. However, because of the  proximity effects of the neighboring phases, it is difficult to make  definite conclusions on the extent of the nematic region based solely on the results of Fig.~\ref{fig:j20.7scan}(b), besides the fact that it is much narrower than suggested na\"{i}vely in Fig.~\ref{fig:pairing}(a).

Thus, we carry out the fixed-parameter, $20\!\times\!8$ cluster calculation as in Fig.~\ref{fig:nematic} for several values of $h$ along the path of the scan in Fig.~\ref{fig:j20.7scan}(b). The results for three such fields, 0.9, 0.96, and 1.0,  are shown in Figs.~\ref{fig:j20.7scan}(c)-(e). Fig.~\ref{fig:j20.7scan}(d) mirrors Fig.~\ref{fig:nematic}(c), clearly placing $h\!=\!0.96$  in the nematic region. The finite-size scaling of the  nematic order  shows little change \cite{sm}, indicating the near-2D character of our results. The $h\!=\!1.0$ point in Fig.~\ref{fig:j20.7scan}(e) shows  saturated ordered moment and a decay of both pair and spin-canting away from the  boundaries, confirming a polarized FM state. The  $h\!=\!0.9$ point in Fig.~\ref{fig:j20.7scan}(c) demonstrates a strong presence of both  dipolar and quadrupolar orders---a sign of the AFM phase.  For all the ($J_2$, $h$) data points contributing to the phase diagram in Fig.~\ref{fig:phasediag}(d), we performed the same type of analysis. 

\begin{figure}[t]
\centering    
\includegraphics[width=1.0\columnwidth]{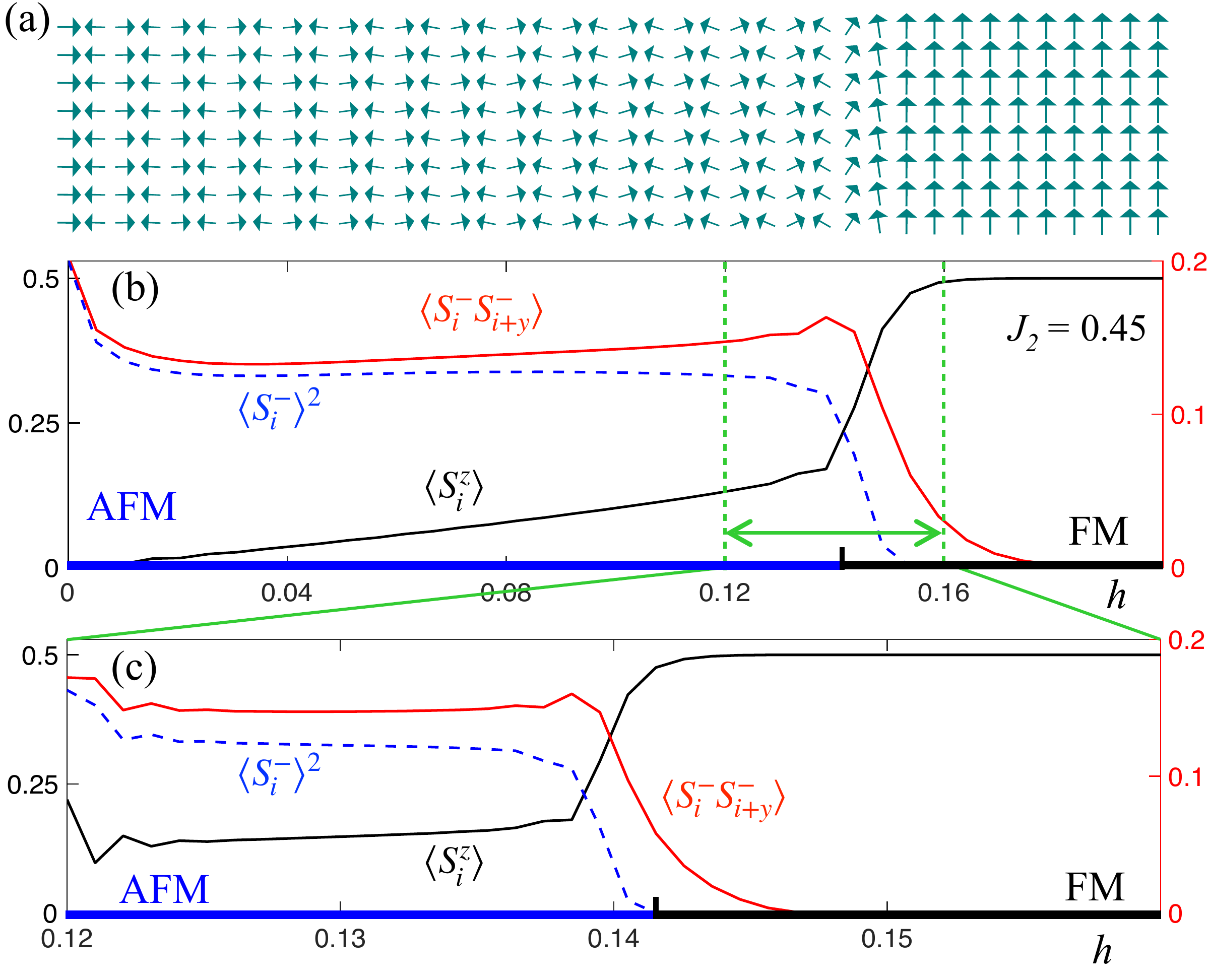}
\vskip -0.35cm
\caption{(a) and (b) Same as (a) and (b) in Fig.~\ref{fig:j20.7scan} for $J_2\!=\!0.45$ and $h$ from 0.0 to 0.2. (c) Same as (b) for $h$ from 0.12 to 0.16.}
\label{fig:j20.45scan}
\vskip -0.6cm
\end{figure}

In Figure~\ref{fig:j20.45scan}, we present the results of the same analysis for $J_2\!=\!0.45$, with the scan in $h$ from 0.0 to 0.2. Unlike the case of Figure~\ref{fig:j20.7scan}, where the  evolution of magnetization suggests second-order transitions, in Fig.~\ref{fig:j20.45scan}(a) and \ref{fig:j20.45scan}(b) one can notice that the canting of  spins  changes to a fully polarized state rather drastically. The transition is at about $h\!\approx\!0.14$, which is also noticeably {\it higher} than the pair-BEC value of $h_{c2}\!=\!0.12$  from Fig.~\ref{fig:pairing}(a). Another feature  is the ``scale-invariance'' of the scan, demonstrated in Fig.~\ref{fig:j20.45scan}(c) by zooming on the narrow field range of 0.12 to 0.16, suggesting the first-order character of the transition. The fixed-parameter calculations described above also find no nematic region between the AFM and FM states, supporting our scenario that  pair attraction leads to a first-order collapse of the multi-pair state directly into the dipolar instead of the nematic phase, in a broad agreement with the proposal of Ref.~\cite{ueda_phasesep}.

The AFM-FM transition remains first-order down to zero field with the boundary shifting to $J_2\!\approx\!0.39$ from the pair-BEC value of $J_2\!\approx\!0.408$, see Fig.~\ref{fig:pairing}(b),  in agreement with $J_2\!=\!0.394$  from the earlier study \cite{edzerofield}. 

{\it Multi-pair states.}---For $J_2\!\alt\!0.6$ (left of the cyan circle in Fig.~\ref{fig:phasediag}), spin-flip pairs attract each other and can form multi-pair states. As a result, the actual transition from the FM phase is above $h_{c2}$ and is into the condensates of these multi-pair states. Furthermore, the quadrupolar nematic phase also extends above the $h_{c2}$ line, see Figs.~\ref{fig:phasediag}(d) and \ref{fig:pairing}(b), for the same reason the dipolar AFM phase is pulled up above the $h_{c1}$ line. 

In the regime associated with the pair-attraction, we  identified condensations from the FM phase into the states with four, six, and eight magnons in a $16\!\times\!8$ cluster, see Ref.~\cite{sm}. They form a devil's staircase of diminishing ranges of $J_2$ before reaching the first-order transition point at  $J_2\!\approx\!0.5$, bearing a resemblance to the results of Refs.~\cite{oleg1,Lauchli09}.
However, an unambiguous confirmation of the higher-multipolar orders associated with the multi-pair BECs is beyond the present study because of the finite-size effects and weak higher-order pairing.

{\it Summary.}---We have established the actual extent of the $d$-wave nematic phase in the phase diagram of the  paradigmatic $J_1$--$J_2$ model using analytical and DMRG insights. The nature of the $d$-wave pairing  is explained and the criteria for the existence of the pair-BEC are elucidated. The sequence of the multi-pair BEC transitions is suggested to bridge the $d$-wave pair-BEC and the first-order FM-AFM  transition lines.  

The nematic state is not stable at zero field  and in the $J_2$ region close to the FM-AFM border because repulsive pair-pair interactions are generally required to ensure finite stiffness of the pair-BEC state. A suppression of the single-spin-flip gap by an attraction to the pair-condensate is shown to lead to a dramatic order-of-magnitude contraction of the nematic phase compared to the na\"{i}ve expectations. The hallmark of the remaining nematic region is the significant drop in the magnetization in a very narrow field range near saturation without any dipolar order. Our work provides vital guidance to the ongoing theoretical and experimental searches of the elusive quantum spin-nematics, arming them with realistic expectations. The proposed scenario and the phase diagram can be expected to be valid for a wide variety of models and materials.

\begin{acknowledgments}
\emph{Acknowledgments.}---%
The work of S.~J. and S.~R.~W. was supported by the NSF through grant DMR-2110041.
The work of J.~R.  was supported by the NSF through grant DMR-2142554.
The work of M.~E.~Z. was supported by ANR, France, Grant No. ANR-15-CE30-0004.
The work of A.~L.~C. was supported by the U.S. Department of Energy, 
Office of Science, Basic Energy Sciences under Award No. DE-SC0021221.
\end{acknowledgments}

\bibliography{ref}

\end{document}


\title{Where is the Quantum Spin Nematic?: Supplemental Material}

\begin{CJK*}{UTF8}{}
\author{Shengtao Jiang  (\CJKfamily{gbsn}蒋晟韬)}
\affiliation{Department of Physics and Astronomy, University of California, Irvine, California 92697, USA}

\author{Judit Romh\'anyi}
\affiliation{Department of Physics and Astronomy, University of California, Irvine, California 92697, USA}

\author{Steven R. White}
\affiliation{Department of Physics and Astronomy, University of California, Irvine, California 92697, USA}

\author{M. E. Zhitomirsky}
\affiliation{Universit\'e Grenoble Alpes, CNRS, LPMMC, 38000 Grenoble, France}
\affiliation{Institut Laue-Langevin, 71 Avenue des Martyrs, CS 20156, 38042 Grenoble Cedex 9, France}

\author{A. L. Chernyshev}
\affiliation{Department of Physics and Astronomy, University of California, Irvine, California 92697, USA}

\date{\today}
\maketitle
\end{CJK*}

\section{Multipolar order parameters}

We are interested in quantum spin orders that break full rotational symmetry without displaying dipolar ordering. These phases are characterized by multipole order parameters such as quadrupole, octupole, etc. Higher quantum spins have larger local Hilbert spaces and naturally exhibit higher-multipole degrees of freedom. In general, spins $S$ allows for on-site order parameters, which transform as rank-$k$ tensor operators, with $k$ taking  values of $0, 1,\hdots, 2S$.

A rank-$k$ tensor operator $\boldsymbol{T}^{(k)}$ has $2k+1$ components that satisfy the commutation relations \cite{Tinkham2003Group}
\begin{eqnarray}
\big[S^z,T^{(k)}_q\big]=qT^{(k)}_q\qquad\text{and}\qquad
\big[S^\pm,T^{(k)}_q\big]=\sqrt{k(k+1)-q(q\pm1)} \, \, T^{(k)}_{q\pm 1}\;,
\label{eq:multipole_gen}
\end{eqnarray}
which enable a systematic construction of the multipole order parameters $T^{(k)}_q$ with $q\in\left[-k,k\right]$ for each $k$. 

For example, the dipole operators correspond to the $k=1$ multiplet with $T_1^{(1)}=S^+$, $T_0^{(1)}=S^z$, and $T_{-1}^{(1)}=S^-$, while the quadrupole operators form a rank-2 tensor with
\begin{eqnarray}
T_2^{(2)}=S^+S^+ ,\quad T_1^{(2)}=-S^+S^z-S^zS^+ ,\quad T_0^{(2)}=\sqrt{\frac{2}{3}}\big(3 S^zS^z-S(S+1)\big) , \quad T_{-1}^{(2)}={T_1^{(2)}}^\dagger ,\ \ \text{and}\ \
T_{-2}^{(2)}={T_2^{(2)}}^\dagger .
\label{eq:quadrupoles}
\end{eqnarray}
Octupolar operators can be generated from $T_3^{(3)}=S^+S^+S^+$ by the repeated application of $\big[S^\pm,T^{(k)}_q\big]$, see Eq.~(\ref{eq:multipole_gen}).

For the $S=\frac 1 2$ systems, quadrupole (nematic) orders require a consideration of the  two-spin operators  in the form of $S^\alpha_1 S^{\beta}_2$ defined on bonds instead of the on-site spin-operators in case of the higher spins. The decomposition of such bond operators consists of the trace, that is the scalar operator $\boldsymbol{S}_1\cdot\boldsymbol{S}_2$, which is SU(2) invariant and commutes with the total spin, the traceless antisymmetric part $\boldsymbol{S}_1\times\boldsymbol{S}_2$, which is a vector (rank-1 tensor), and the traceless symmetric part corresponding to spin-quadrupole operators (rank-2 tensor) $Q^{\alpha\beta}_{12}=S^\alpha_1S^\beta_2+S^\beta_1S^\alpha_2-\delta_{\alpha\beta}\frac 2 3 \big(\boldsymbol{S}_1\cdot\boldsymbol{S}_2\big)$.

The  quadrupole bond operators can be expressed in the time-reversal invariant basis  
\begin{align}
&\left|s\right>=\frac{1}{\sqrt{2}}\big(\left|\uparrow\downarrow\right>-\left|\downarrow\uparrow\right>\big),\qquad
\left|t_x\right>=\frac{i}{\sqrt{2}}\big(\left|\uparrow\uparrow\right>-\left|\downarrow\downarrow\right>\big),\qquad
\left|t_y\right>=\frac{1}{\sqrt{2}}\big(\left|\uparrow\uparrow\right>+\left|\downarrow\downarrow\right>\big),\qquad
\left|t_z\right>=\frac{-i}{\sqrt{2}}\big(\left|\uparrow\downarrow\right>+\left|\downarrow\uparrow\right>\big),
\label{eq:bond_basis}
\end{align}
as $Q^{\alpha\beta}_{12}=-\frac 1 2 \big(|t_\alpha\rangle\langle t_\beta|+|t_\beta\rangle\langle t_\alpha|\big)+\frac 1 3 \delta_{\alpha\beta}\sum_\gamma|t_\gamma\rangle\langle t_\gamma|$.
A bond nematic state can be written as a linear combination of the triplet states $\left|\boldsymbol{t}\right>=\sum_{\alpha} \bar{t}_\alpha \left|t_\alpha\right>$  with real coefficients $\bar{t}_\alpha$. Such a state is time-reversal symmetric, and has no dipole order, $\left<\boldsymbol{t}\right| S_i^\alpha\left|\boldsymbol{t}\right>\equiv 0$, as only the quadrupole matrix elements are finite.

While the expectation values of dipolar operators are zero in a purely quadrupolar state, the reverse is not true; quadrupolar order parameters can be finite in a magnetic state. For example, the magnetic state $\left|\uparrow\uparrow\right>\!=\!\frac{1}{\sqrt{2}}\big(\left|t_y\right>-i \left|t_x\right>\big)$ gives $\left<S^z_{\rm{tot}}\right>=1$ and at the same time $\left<Q^{zz}\right>=\frac 1 3$.

Bond-nematic phases in spin-half frustrated magnets can arise from a condensation of bound magnon pairs in the field-polarized paramagnet. The two-magnon state is directly related to the quadrupole operator ${T_{-2}^{(2)}}_{ij}=-S^-_iS^-_j$ 
\begin{equation}
\sum_{i,j} \psi_{ij} {T_{-2}^{(2)}}_{ij} |\uparrow\uparrow\hdots\uparrow\rangle=\sum_{i,j} \psi_{ij} \big(Q^{x^2-y^2}_{ij}- i Q^{xy}_{ij}\big) |\uparrow\uparrow\hdots\uparrow\rangle\;,
\end{equation}
with the bound state in the $Q^{x^2-y^2}_{ij}$ and $Q^{xy}_{ij}$ channels. The orbital (lattice) symmetry of the bound state is encoded in the $\psi_{ij}$ coefficients that are determined from the solution of the two-magnon Schroedinger equation (SE) considered in the next Section. While the nomenclature of the $Q^{\alpha\beta}_{ij}$ operators is reminiscent of that of the $L=2$ orbital states, it is unrelated to the $d$-wave nature of the bound states discussed below.

The ordering of the dipole and quadrupole moments can be expressed via the spin and quadrupole structure factors,  ${\cal S}^{\alpha\beta}({\bf k})=\langle {S}^\alpha_{\bf k}\, {S}^\beta_{\bf -k}\rangle$ and ${\cal Q}^{\alpha\beta}({\bf k})=\langle {Q}^{\alpha\beta}_{\bf k}\, {Q}^{\alpha\beta}_{\bf -k}\rangle$, respectively.
One expects finite Bragg peaks in both ${\cal S}({\bf k})$ and ${\cal Q}({\bf k})$ in the conventional magnetic phase, whereas in the nematic state Bragg peaks are present only in  ${\cal Q}({\bf k})$.

A general quadrupole-quadrupole correlation function has the form 
$\langle Q^{\alpha\beta}_{i,j}Q^{\alpha\beta}_{i+r,j+r^\prime}\rangle$,
which can be measured numerically in finite clusters. However, for the symmetry-broken states, the expectation value of the nematic order parameter  $\langle Q^{\alpha\beta}_{ij}\rangle$ can be also detected directly, similarly to the ordered moment $\langle S^z_i \rangle$ in the dipolar magnetic states. This is used in our DMRG approach, see Secs.~\ref{j1scan} below and the main text.

\section{Magnon bound states}
\label{MBS}

\subsection{Magnon interaction}
\label{potential}

We restrict ourselves to the bound-state problem in a fully polarized FM phase  of an isotropic spin-$S$ Heisenberg model on a Bravais lattice in a field
\begin{align}
{\cal H}=\frac12\sum_{i,{\bm \delta}}J_{\bm \delta}\,{\bf S}_i\cdot {\bf S}_j-H\sum_i S^z_i, 
\label{HH}
\end{align}
where  ${\bm \delta}\!=\!{\bf r}_j-{\bf r}_i$ spans all non-zero $n$th-neighbor couplings that are present in the model.

The kinetic energy of a magnon, or a single-spin-flip eigenstate $|\psi_{1,{\bf q}}\rangle\!=\!1/\sqrt{N}\sum_i e^{-i{\bf q}{\bf r}_i}S^-_i|0\rangle$, relative to the ground-state energy $E_0$ of $|0\rangle\!=\!|\uparrow\uparrow\uparrow\dots\rangle$, can be written as \cite{zhitomirsky2010}
\begin{align}
\varepsilon_{{\bf q}}=H+S\big({\cal J}_{\bf q}-{\cal J}_{0}\big) , \ \ \ \ \mbox{with} \ \ \ \ {\cal J}_{\bf q}=\sum_{\bm \delta} J_{\bm \delta}e^{i {\bf q}{\bm \delta}},
\label{Ek}
\end{align}
where ${\cal J}_{\bf q}\!=\!{\cal J}_{-\bf q}$ as $J_{\bm \delta}\!=\!J_{-\bm \delta}$ on a Bravais lattice. Assuming that $\varepsilon_{{\bf q}}$ has an absolute minimum at the momentum ${\bf Q}$ and given the simple Zeeman-like dependence of energy on the field, one can find the critical field for a transition to a single-spin-flip BEC   from the condition $\varepsilon_{{\bf Q}}\!=\!0$, which yields
\begin{align}
H_{s1}=S\big({\cal J}_{0}-{\cal J}_{\bf Q}\big) .
\label{Hs1}
\end{align}
In the absence of an attraction between magnons, the expression (\ref{Hs1})  is valid for any value of $S$. Thus, one refers to $H_{s1}$ and to the single-magnon BEC transitions itself as the the ``classical'' ones in this case. 

Magnon interaction requires a consideration of the two-spin-flip wavefunction, whose generic form is given by
\begin{align}
|\psi_{2}\rangle=\frac12\sum_{i,j}  \psi_{i,j} S^-_i S^-_j|0\rangle.
\label{psi2}
\end{align}
The reason why this function cannot be built directly from the basis of single-spin-flip {\it operators}, which are complete and orthogonal in the single-particle sector, is the non-commutativity of spin operators, the problem recognized early on \cite{Bethe31,FeynmanSM,Dyson56}. The commonly followed resolution \cite{Wortis_63,Mattis} is to operate directly in the basis of the $|\psi_{2}\rangle$ wavefunctions (\ref{psi2}), which is complete, but not orthogonal. Despite this latter inconvenience, the {\it exact} two-particle Schroedinger equation (SE) can be obtained  by standard, if tedious, manipulations \cite{zhitomirsky2010} to yield 
\begin{align}
(E-\varepsilon_{\frac{\bf K}{2}+{\bf q}}-\varepsilon_{\frac{\bf K}{2}-{\bf q}})\psi_{\bf K}({\bf q})
=\frac{1}{2N}\sum_{\bf p} V_{\bf K}({\bf q},{\bf p})\psi_{\bf K}({\bf p}), \ \ \ \mbox{with} \ \ \ \psi_{\bf K}({\bf q})=\frac{1}{N}\sum_{i,j}e^{-i\frac{\bf K}{2}({\bf r}_i+{\bf r}_j)} e^{-i{\bf q}({\bf r}_j-{\bf r}_i)} \psi_{i,j},
\label{2MagSE}
\end{align}
and the magnon interaction potential given by 
\begin{align}
V_{\bf K}({\bf q},{\bf p}) = {\cal J}_{{\bf q}-{\bf p}}+{\cal J}_{{\bf q}+{\bf p}}-{\cal J}_{\frac{{\bf K}}{2}-{\bf q}}-{\cal J}_{\frac{\bf K}{2}+{\bf q}}.
\label{2MagVKqp}
\end{align}
The first two terms in $V_{\bf K}({\bf q},{\bf p})$ in (\ref{2MagVKqp}) adhere to the conventional ``potential-like'' form, as they depend  on the momentum transfer between  initial and final particles in the scattering, with the momenta ${\bf K}/2\pm{\bf q}$ and ${\bf K}/2\pm{\bf p}$, respectively. The last two terms depend on the individual  particle momenta  and originate from the ``hard-core'' repulsion of spin flips.

Given the single-magnon energy in (\ref{Ek}) and magnon interaction potential in (\ref{2MagVKqp}), it is made explicit by SE (\ref{2MagSE}) that the magnon interaction is  manifestly a $1/S$ effect, with the role of pairing diminishing in the large-$S$ limit and single-magnon BEC retaining its  classical value. 

We also would like to remark that an expression {\it identical} to (\ref{2MagVKqp}) is trivially obtained in the $O(S^0)$ order of expansion of (\ref{HH}) within the perfectly commutative and orthogonal {\it bosonic} basis  of the standard $1/S$-approximation, obviously leading to a form of the two-magnon SE that is identical to (\ref{2MagSE}).

\subsection{Simple case, Cooper problem analogy, and higher harmonics}
\label{simple}

It is instructive to have a general intuition on solving SE in (\ref{2MagSE}) for a simple case. Consider a continuum limit for particles of mass $m$ and energies $\varepsilon_{\bf k}\!=\!{\bf k}^2/2m$ that are attracted via a  $\delta$-functional potential in real space, $V({\bf r}_1,{\bf r}_2)\!=\! -\alpha\delta({\bf r}_1-{\bf r}_2)$, which corresponds to a constant in the momentum space, $V_{\bf K}({\bf q},{\bf p})\!=\!-\alpha$. This reduces  (\ref{2MagSE}) to an algebraic equation for $\psi$ that is trivially solved  as 
\begin{align}
\psi_{\bf K}({\bf q})=-\frac{\alpha}{2}\cdot\frac{C_{\bf K}}{E-E_{\bf K}-{\bf q}^2/m},\ \ \ \ \mbox{with} \ \ \ \ C_{\bf K}=\frac{1}{N}\sum_{\bf p} \psi_{\bf K}({\bf p}),
\label{SE_alpha}
\end{align}
where $E_{\bf K}\!=\!{\bf K}^2/4m$ is the bottom of the two-particle continuum for the total momentum ${\bf K}$. The second equation in (\ref{SE_alpha}) provides a self-consistency condition, which is also an implicit equation on the pair's energy   $E$
\begin{align}
1=-\frac{\alpha}{2N}\sum_{\bf p}\frac{1}{E-E_{\bf K}-{\bf p}^2/m}.
\label{SE_E}
\end{align}
Finding $E\!<\!E_{\bf K}$ from (\ref{SE_E}) constitutes a solution of the bound-state problem. This  consideration is analogous to a textbook problem of the bound state for a single particle in a free space with a  $\delta$-functional potential well  \cite{Galitskii}, also demonstrating a mapping of the two- to  one-body problem. A direct integration in Eq.~(\ref{SE_E}) in the $D$-dimensional continuum demonstrates the existence of a bound state  for an {\it arbitrarily weak} attraction in the dimensions $D\!=\!1$ and $D\!=\!2$, and requires $\alpha$ to exceed a threshold value of order $1/m$ in $D\!=\!3$ \cite{Galitskii}. Since we are interested in the case of $D\!=\!2$, the solution for it is 
\begin{align}
E-E_{\bf K}= \frac{\Lambda}{m}\cdot e^{-\frac{8\pi}{\alpha m}},
\label{SE_2D}
\end{align}
where $\Lambda$ is a large-momentum cut-off and the non-analytic dependence on the coupling constant $\alpha$ is clear. These results rely exclusively on the density of states of the non-relativistic particles in $D$-dimensions and are expected to hold for {\it any} short-range attractive potential in the continuum \cite{Galitskii}. One of the well-known realizations of such conditions is the celebrated Cooper problem of the short-range attraction of electrons near the Fermi surface \cite{Annett}. The presence of the finite Fermi-momentum cutoff leads to an effective {\it dimensional reduction} from 3D to 2D, resulting in the 2D-like density of states and the pairing gap that closely follows  Eq.~(\ref{SE_2D}).

For a realization of the pair-BEC, the lowest bound-state energy needs to be below the {\it absolute minimum} of the two-particle continuum, which typically occurs at a high-symmetry ${\bf K}$-point. In an example above, this is the ${\bf K}\!=\!0$ point. There, a systematic symmetry consideration of the pairing in the higher partial-wave channels is also possible. Given the mapping onto the one-body problem in the continuum, it can be made explicit that with an exception of the $s$-wave channel, which is exposed above, all other channels have nodes of their corresponding pairing potential that pass through the minimum of the single-particle energy. This crucial feature leads to an effective  {\it dimensional increase}, as opposed to the dimensional reduction, suggesting that the weak attraction in the continuum in 2D can only create $s$-wave bound states, but not the ones in the higher-harmonics, which would require the corresponding coupling strengths to exceed threshold values of order $1/m$.

\subsection{Pairing in the $J_1$--$J_2$ Heisenberg model}
\label{J1J2pairing}

We now turn to the problem of pairing in the ferro-antiferromagnetic  $S=1/2$ $J_1$--$J_2$ Heisenberg model on the square lattice with the ferromagnetic $J_1\!=\!-1$ taken as a unit of energy and antiferromagnetic $J_2\!>\!0$. The convenient shorthand notation ${\cal J}_{\bf q}$ in (\ref{Ek}) can be rewritten using standard hopping amplitudes 
\begin{align}
{\cal J}_{\bf q}=-4\gamma_{\bf q}+4J_2\gamma^{(2)}_{\bf q}, \ \ \ \mbox{with} \ \ \ \gamma_{\bf q}=\frac12\big(\cos q_x+\cos q_y\big), \ \ \ \mbox{and} \ \ \ \gamma^{(2)}_{\bf q}=\cos q_x\cos q_y.
\label{Jq}
\end{align}
The single-magnon energy (\ref{Ek}) for $S=1/2$ is given by  
\begin{align}
\varepsilon_{{\bf q}}=H+2\big(1-\gamma_{\bf q}\big)-2J_2\big(1-\gamma^{(2)}_{\bf q}\big)
=\varepsilon_{{\bf Q}}- 2\gamma_{\bf q}+2J_2\big(1+\gamma^{(2)}_{\bf q}\big),
\label{EkJ1J2}
\end{align}
where we use that for $J_2\!>\!0.5$, the minima of $\varepsilon_{{\bf q}}$ are at the non-trivial ${\bf Q}\!=\!(0,\pi)[(\pi,0)]$ points, with $\varepsilon_{{\bf Q}}\!=\!H+2-4J_2$, leading to $H_{s1}\!=\!-2+4J_2$, see (\ref{Hs1}).  This is the range of $J_2$ that we will consider below, although extension to the case of $J_2\!<\!0.5$, for which the minimum of $\varepsilon_{{\bf q}}$ is at ${\bf Q}\!=\!0$, is easily made. 

In addition to the hopping amplitude $\gamma_{\bf q}$ in (\ref{Jq}), one can see an immediate utility in introducing other combinations of lattice harmonics  of different symmetry
\begin{align}
\gamma^{-}_{\bf q}=\frac12\big(\cos q_x-\cos q_y\big), \ \ \ \gamma^{s_{xy}}_{\bf q}=[\gamma^{(2)}_{\bf q}=]\cos q_x\cos q_y, \ \  \ \mbox{and} \ \ \ \gamma^{d_{xy}}_{\bf q}=\sin q_x\sin q_y,
\label{gammas}
\end{align}
as they permit an easy symmetrization of the single-magnon energies in SE (\ref{2MagSE}) and a separation of interaction potential (\ref{2MagVKqp}) into the partial harmonics using 
\begin{align}
\gamma_{{\bf q}-{\bf p}}+\gamma_{{\bf q}-{\bf p}}=2\gamma_{\bf p}\gamma_{\bf q}+2\gamma^-_{\bf p}\gamma^-_{\bf q}, \ \ \ \ \ \  \gamma^{(2)}_{{\bf q}-{\bf p}}+\gamma^{(2)}_{{\bf q}-{\bf p}} =2\gamma^{s_{xy}}_{\bf p}\gamma^{s_{xy}}_{\bf q}+2\gamma^{d_{xy}}_{\bf p}\gamma^{d_{xy}}_{\bf q}.
\label{gammas_sep}
\end{align}
Introducing binding energy of the pair $2\Delta\!>\!0$ relative to the absolute minimum of the two-magnon continuum via $E\!=\!-2\Delta+2\varepsilon_{{\bf Q}}$, converts the energy difference in the left-hand side of SE in (\ref{2MagSE}) to
\begin{align}
E-\varepsilon_{\frac{\bf K}{2}+{\bf q}}-\varepsilon_{\frac{\bf K}{2}-{\bf q}} = 
-2\Delta+4\big(\gamma_{\frac{\bf K}{2}}\gamma_{\bf q}+\gamma^-_{\frac{\bf K}{2}}\gamma^-_{\bf q}\big)
-4J_2\big(1+\gamma^{s_{xy}}_{\frac{\bf K}{2}}\gamma^{s_{xy}}_{\bf q}+\gamma^{d_{xy}}_{\frac{\bf K}{2}}\gamma^{d_{xy}}_{\bf q}\big),
\label{E_sym}
\end{align}
while the interaction potential (\ref{2MagVKqp}) becomes
\begin{align}
V_{\bf K}({\bf q},{\bf p}) = -8\gamma_{\bf q}\big(\gamma_{\bf p}-\gamma_{\frac{\bf K}{2}}\big) -8\gamma^-_{\bf q}\big(\gamma^-_{\bf p}-\gamma^-_{\frac{\bf K}{2}} \big)+
8J_2\gamma^{s_{xy}}_{\bf q}\big(\gamma^{s_{xy}}_{\bf p}-\gamma^{s_{xy}}_{\frac{\bf K}{2}}\big)+8J_2\gamma^{d_{xy}}_{\bf q}\big(\gamma^{d_{xy}}_{\bf p}-\gamma^{d_{xy}}_{\frac{\bf K}{2}}\big),
\label{VKqp_sym}
\end{align}
with the terms clearly assuming the separable partial-wave decomposition structure, although with asymmetric ``shifts'' of the harmonics by $\gamma^{\gamma}_{\frac{\bf K}{2}}$ due to the hard-core components of the original potential (\ref{2MagVKqp}). 

\begin{figure}[t]
\centering    
\includegraphics[width=1\columnwidth]{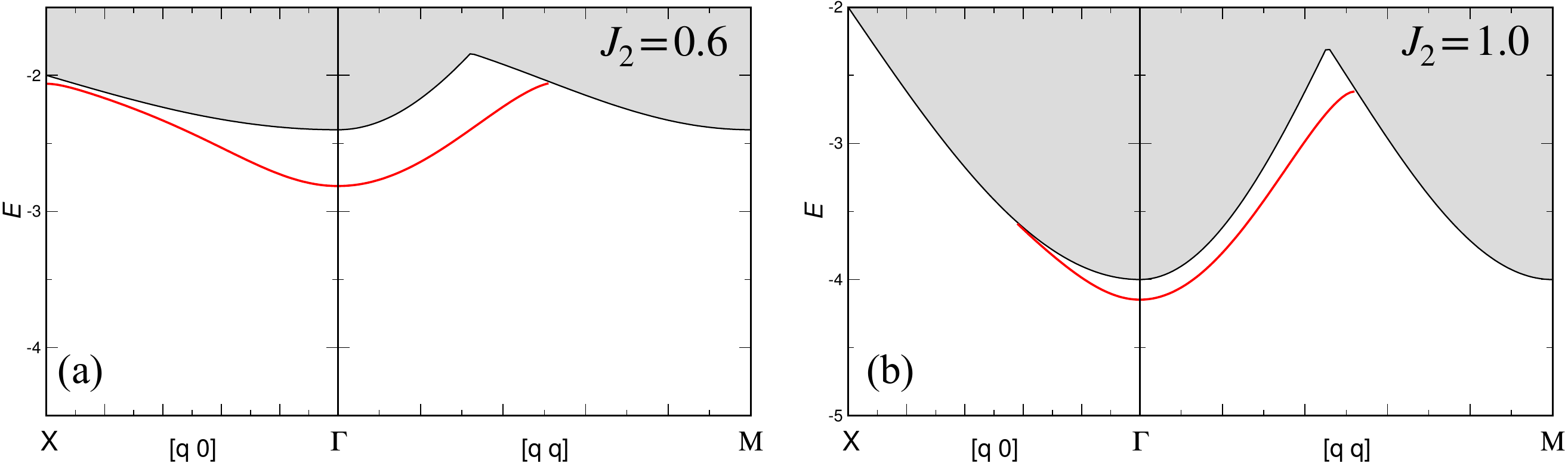}
\caption{The two-magnon continuum (shaded area) and the bound state energies (lines) along the representative ${\bf K}=[q_x,q_y]$ cuts in the Brillouin zone for (a) $J_2=0.6$ and (b) $J_2=1.0$. $M=[\pi,\pi]$ and $X=[\pi,0]$.}
\label{fig:BS_K}
\end{figure}

For the magnon-band minima at ${\bf Q}\!=\!(0,\pi)[(\pi,0)]$ points, there only two distinct total momenta of their pair at which pair-BEC can occur, ${\bf K}\!=\!0$ and ${\bf K}\!=\!(\pi,\pi)$. While rigorous arguments can be given to prove that the ${\bf K}\!=\!(\pi,\pi)$ sector does not support pairing in any orbital channel in the  range of $J_2\!>\!1/\pi$, we skip them in favor of the visual demonstration of the numerical solution for the bound state problem (\ref{2MagSE}) with the energy denominator (\ref{E_sym}) and pairing potential (\ref{VKqp_sym}) for ${\bf K}$ along a representative cut of the Brillouin zone and for two representative $J_2=0.6$ and $J_2=1.0$, see Fig.~\ref{fig:BS_K}. This Figure clearly points to the importance of the ${\bf K}\!=\!0$ sector in our case, considered next.

\subsubsection{${\bf K}\!=\!0$ sector}

For the  ${\bf K}\!=\!0$ sector,  $\gamma_{\frac{\bf K}{2}}\!=\!\gamma^{s_{xy}}_{\frac{\bf K}{2}}\!=\!1$ and $\gamma^-_{\frac{\bf K}{2}}\!=\!\gamma^{d_{xy}}_{\frac{\bf K}{2}}\!=\!0$, converting the energy part of SE (\ref{E_sym}) to 
\begin{align}
\Delta E_{0}({\bf q})=\big(E-\varepsilon_{\frac{\bf K}{2}+{\bf q}}-\varepsilon_{\frac{\bf K}{2}-{\bf q}}\big)\big|_{{\bf K}=0} = -2\Delta-4J_2\big(1+\gamma^{(2)}_{\bf q}\big)+4\gamma_{\bf q} = -2\Delta-2\big(\varepsilon_{\bf q}-\varepsilon_{\bf Q}\big)<0,
\label{E_sym0}
\end{align}
where the binding energy $\Delta$ is relative to the $2\varepsilon_{\bf Q}$ minimum,  
and the interaction potential (\ref{VKqp_sym}) becomes, 
\begin{align}
V_{0}({\bf q},{\bf p}) = -8\gamma_{\bf q}\big(\gamma_{\bf p}-1\big)-8\gamma^-_{\bf q}\gamma^-_{\bf p}+
8J_2\gamma^{s_{xy}}_{\bf q}\big(\gamma^{s_{xy}}_{\bf p}-1\big)+8J_2\gamma^{d_{xy}}_{\bf q}\gamma^{d_{xy}}_{\bf p}=\sum_\gamma V_\gamma R_\gamma({\bf q})\widetilde{R}_\gamma ({\bf p}),
\label{VKqp_sym0}
\end{align}
with  $\gamma\!=\!\{s,d,s_{xy},d_{xy}\}$,  $V_\gamma\!=\!\{8,8,-8J_2,-8J_2\}$, $R_\gamma({\bf q})\!=\!\{\gamma_{\bf q}, \gamma^-_{\bf q}, \gamma^{s_{xy}}_{\bf q}, \gamma^{d_{xy}}_{\bf q}\}$,  and $\widetilde{R}_\gamma({\bf p})\!=\!\{\gamma_{\bf p}-1, \gamma^-_{\bf p}, \gamma^{s_{xy}}_{\bf p}-1, \gamma^{d_{xy}}_{\bf p}\}$.

While it is tempting to conclude that this form of interaction already guarantees   decomposition of the SE in (\ref{2MagSE}) into a fully orthogonal set of  algebraic equations on pairing in separate channels, the  analysis provided so far is based entirely on the lattice-harmonic expansion of the interaction potential and magnon energies. Continuing without any symmetry analysis for a moment, the SE  (\ref{2MagSE}) for the ${\bf K}\!=\!0$ sector can be rewritten as 
\begin{align}
\psi_{0}({\bf q})
=-\frac12\sum_\gamma V_\gamma R_\gamma({\bf q})\cdot\frac{C_{\gamma}}{\Delta E_{0}({\bf q})},\ \ \ \ \mbox{with} \ \ \ \ C_\gamma=\frac{1}{N}\sum_{\bf p} \widetilde{R}_\gamma({\bf p})\psi_{0}({\bf p}).
\label{SE_0}
\end{align}
Although it is clear that the SE is, indeed, reduced to a set of algebraic equations for $\psi_{0}({\bf q})$,  these equations are, or can be, coupled, as the constants  $C_\gamma$ in (\ref{SE_0}) can have contributions from the $\gamma'\!\neq\!\gamma$ components of $\psi_{0}({\bf q})$. 

The partial waves in (\ref{VKqp_sym0}) and (\ref{SE_0}) are of the $s$,  $s_{xy}$, $d_{x^2-y^2}$, and $d_{xy}$ character, with the mirror-plane symmetries at the ${\bf K}\!=\!0$ point separating $d_{x^2-y^2}$ and $d_{xy}$ partial-wave solutions of the SE, while the two generalized $s$-wave-like solutions are allowed to mix. Thus,  Eq.~(\ref{SE_0}) for the  ${\bf K}\!=\!0$ sector decouples into two independent algebraic equations for the $d$-waves and two coupled equations for the $s$-waves.

The resultant  equations on the pairing energy for  the $d_{x^2-y^2}$ ($d$ for brevity) and $d_{xy}$ waves are given by
\begin{align}
d: \ \ \ \ \ 1=-\frac{4}{N}\sum_{\bf p}\frac{(\gamma^-_{\bf p})^2}{\Delta E_{0}({\bf p})}, \ \ \ \ \ \  \ \ \ \mbox{and} \ \ \   
d_{xy}: \ \ \ \ \ 1=\frac{4J_2}{N}\sum_{\bf p}\frac{(\gamma^{d_{xy}}_{\bf p})^2}{\Delta E_{0}({\bf p})}, 
\label{Ed_0}
\end{align}
of which the $d_{xy}$-wave does not permit any solutions having negatively-defined right-hand side.
The $d$-wave equation yields a solution of our ultimate interest that is discussed in more detail below. 

The coupled $s$-wave equations can be written as
\begin{align}
&C_s=C_s I_{11}+C_{s_{xy}} I_{12}, \ \ \ \mbox{and} \ \ \ C_{s_{xy}}=C_{s_{xy}}  I_{22}+C_s I_{21}, \ \ \ \mbox{with} \ \ \ I_{11}=-\frac{4}{N}\sum_{\bf p}\frac{\gamma_{\bf p}(\gamma_{\bf p}-1)}{\Delta E_{0}({\bf p})},
\label{Es_0} \\
&  I_{22}=\frac{4J_2}{N}\sum_{\bf p}\frac{\gamma^{s_{xy}}_{\bf p}(\gamma^{s_{xy}}_{\bf p}-1)}{\Delta E_{0}({\bf p})}, \ \ \ 
I_{12}=\frac{4J_2}{N}\sum_{\bf p}\frac{\gamma^{s_{xy}}_{\bf p}(\gamma_{\bf p}-1)}{\Delta E_{0}({\bf p})}, \ \ \ 
I_{21}=-\frac{4}{N}\sum_{\bf p}\frac{\gamma_{\bf p}(\gamma^{s_{xy}}_{\bf p}-1)}{\Delta E_{0}({\bf p})}.
\label{Es_0a}
\end{align}
While, ultimately, one can simply use these expressions to numerically prove that they are unable to yield a viable solution for the binding energy, having an intuition on ``why'' would be very useful. For that, let us consider only the diagonal $I_{\gamma\gamma}$ terms, because if they are unable to sustain a solution on their own, the situation is unlikely to improve by the coupling between the two channels. 

For the $I_{22}$ in the $s_{xy}$-wave channel, the situation is similar to the $d_{xy}$-wave above in that the coupling is due to a repulsive $J_2$ term, but it contains a ``hard-core shift'' in the pairing potential, which makes the outcome less certain analytically. However, with a minimal effort, one is able to prove that $I_{22}$ does indeed remain negative and yields no feasible solutions for the binding energy, 

The interaction  in the $s$-wave channel in $I_{11}$ is attractive ($J_1\!=\!-1\!<\!0$), but it has a hard-core shift in the pairing potential, which is repulsive and is often quoted as the reason for the suppression of the $s$-wave pairing. While this is  a valid mechanism for the lack of pairing in the pure ferromagnetic cases, which never result in the bound states that occur below the absolute minimum of the continuum, this is not the reason why the $s$-wave is projected out of a pairing in our case. A counterexample is always useful. If we ignore the momentum-dependence in $\Delta E_{0}({\bf q})$ in the denominator of the integral in $I_{11}$ in (\ref{Es_0}) for a moment (flat magnon bands), it will yield a solution for the bound-state energy that is actually {\it degenerate} with the one for the $d$-wave in (\ref{Ed_0}). Yet, for the true $\Delta E_{0}({\bf q})$ from (\ref{E_sym0}), the $s$-wave solution is absent. Hence, the lack of the binding is elsewhere. 

The resolution is that the extended $s$-wave pairing harmonic $\gamma_{\bf p}\!\propto\!(\cos q_x+\cos q_y)$ in (\ref{Es_0}) has nodes of zeros passing through  the energy minima of the denominator $\Delta E_{0}({\bf p})$, which are at the magnon energy minima ${\bf Q}\!=\!(0,\pi)[(\pi,0)]$, leading to a suppressed attraction via the ``dimensional increase'' mechanism and giving no bound-state solution.
For the smaller values of $J_2\!<\!0.5$, the magnon band minima migrate back to the ferromagnetic minima at ${\bf Q}\!=\!0$, where the ``hard-core" reasoning takes over in forbidding the ${\bf K}\!=\!0$ bound state in the $s$-wave channel. 

Altogether, the $d$-wave ${\bf K}\!=\!0$ magnon bound-state in (\ref{Ed_0}) is the only candidate to have a BEC in the  ferro-antiferromagnetic $J_1$--$J_2$ Heisenberg model on the square lattice. We consider it in some more details next.

\subsubsection{$D$-wave solution and asymptotics}

We rewrite the equation on the $d$-wave pairing energy (\ref{Ed_0}) in two different ways
\begin{align}
1=\frac{1}{2N}\sum_{\bf p}\frac{\big(\cos p_x-\cos p_y\big)^2}{\Delta+2J_2\big(1+\cos p_x\cos p_y\big)-\big(\cos p_x+\cos p_y\big)}, \ \ \ \ \mbox{and} \ \ 1=-\frac{V_d}{2N}\sum_{\bf p}\frac{\big(\gamma^-_{\bf p}\big)^2}{-2\Delta-2\big(\varepsilon_{\bf p}-\varepsilon_{\bf Q}\big)},
\label{Ed_1}
\end{align}
with the first one being handy to use in the numerical integration that results in the curves for $\Delta$ vs $J_2$ in Fig.~2(a) and $h_{c2}\!=\!h_{c1}+\Delta$  in Fig.~1(a) of the main text. The second one, with $V_d\!=\!8$ in units of $|J_1|$ as before, is useful to relate back to the very beginning of our bound-state consideration, the $s$-wave pairing in the continuum. Taking the $J_2\!\gg\!1$ limit for the denominator in (\ref{Ed_1}) and expanding near  the magnon energy minima ${\bf Q}\!=\!(0,\pi)[(\pi,0)]$ yields $\Delta E_{0}({\bf p})\!=\!-2\Delta-{\bf p}^2/m_{\bf Q}$ with $1/m_{\bf Q}\!=\!2J_2$.

\begin{figure}[t]
\centering    
\includegraphics[width=1\columnwidth]{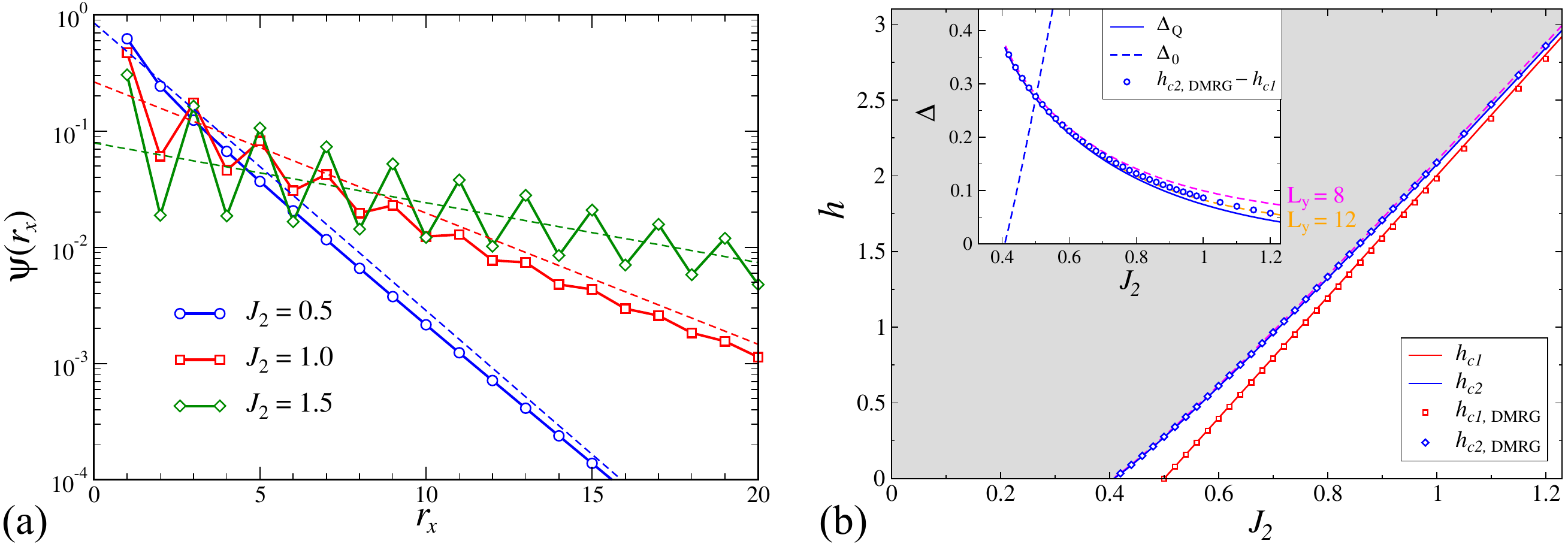}
\caption{(a) Numerical results for the $d$-wave bound-state wave-function  $\psi_d({\bf r}_x)$ on a semi-log plot for three representative values of $J_2$ together with the asymptotic result $\psi_d({\bf r})\!=\! Ae^{-|{\bf r}|/\xi}$, with the correlation length $\xi\!=\!0.8 e^{\pi J_2/2}$ for all three sets. (b)  Same as in Fig.~1(a) and Fig.~2(a) of the main text, with the analytical results for $\Delta$ in (\ref{Ed_1}) obtained using integration on an $L_x\times L_y$ lattice, with  effective $L_x=\infty$ and two values of the width, $L_y=8$ and $L_y=12$.}
\label{fig:FS_Delta}
\end{figure}

Since the zero-nodes of the  pairing harmonic  $\gamma^-_{\bf p}$ are along the $q_x\!=\!\pm q_y$ diagonals, perfectly missing the magnon band minima, the coupling of the low-energy magnons is {\it finite} at ${\bf Q}$, allowing us to approximate $(\gamma^-_{\bf p})^2\!\approx\!1$ near ${\bf Q}$ and giving us all the ingredients of the ``$s$-wave-like'' pairing. Thus, in the weak-coupling limit, integration in (\ref{Ed_1})  yields the answer that mirrors that of Eq.~(\ref{SE_2D})
\begin{align}
2\Delta\approx\frac{\Lambda}{m_{\bf Q}}\cdot e^{-\frac{4\pi}{m_{\bf Q}V_d}}=2J_2 \Lambda \cdot e^{-\pi J_2},
\label{DeltaD}
\end{align}
with $\Lambda$ being the large-momentum cutoff and an extra factor $1/2$ in the exponent originating from the two inequivalent minima of the magnon band at ${\bf Q}\!=\!(0,\pi)$ and $(\pi,0)$. Thus, the pairing in the $d$-wave channel in the considered problem indeed maps onto the textbook $s$-wave solution. One of the key consequence of that is the absence of the upper limit on $J_2$ for such a pairing to occur.  

The other consequence is the exponential decay of the pair wave-function with the distance, $\psi_d({\bf r})\!\sim\! e^{-|{\bf r}|/\xi}$, with the correlation length that can be inferred from the virial theorem, ${\bf q}_\xi^2/2m_{\bf Q}\!\simeq\!\Delta$, as $\xi\!\sim\!e^{\pi J_2/2}$. 
In Figure~\ref{fig:FS_Delta}(a), we show the real-space amplitudes of the bound-state wave-function $\psi_d({\bf r})$ from (\ref{SE_0}) along the line ${\bf r} = (r_x,0)$ for several representative values of $J_2$ together with the asymptotic exponential behavior with $\xi\!=\!0.8e^{\pi J_2/2}$.
Lastly,  for $J_2\!<\!0.5$, the magnon band minimum shifts to ${\bf Q}\!=\!0$, quickly terminating pairing at $J_2^c\!=\!0.4077593304754(1)$, as can be inferred numerically from (\ref{Ed_1}). This is the end-point of the $h_{c2}$ line in Fig.~1(a) in zero field and $\Delta_0\!=\!0$ point in Fig.~2(a) of the main text, see also Fig.~\ref{fig:FS_Delta}(b) and its inset.

One can emulate the finite-size effects of the DMRG results for the paring energy $\Delta$ that can be seen in Fig.~1(a) and Fig.~2(a) of the main text, by taking the analytical integral for $\Delta$ in (\ref{Ed_1}) on an $L_x\times L_y$ lattice. The results of such an effort, with two values of $L_y$ and effectively infinite $L_x$ are shown in Fig.~\ref{fig:FS_Delta}. The results are nearly matched by the width $L_y=12$, not $L_y=8$, which can be interpreted in favor of the open-periodic boundary conditions employed by DMRG as giving a better approximation of the thermodynamic limit than its nominal width would suggest \cite{FS}.

\section{multi-pair states}
\label{multpair}

In this Section, we provide a comparison  of the energies of the states  obtained from the DMRG calculations in the $16\!\times\!8$ cluster with fixed total $S^z$, which gives us the energies of the multi-pair states  as a function of  $J_2$. The energy of the $n$-magnon state is denoted as $E(n)$ and corresponds to the state with $n$ spin flips from the fully polarized ferromagnet. In the following, we consider only even values of $n$, because the states with odd values have consistently higher energies than the nearest even ones, indicating that the gap to single-magnon excitations is always present in the considered range of $J_2\!>\!0.39$ and for the low concentration of spin flips, $n/N_{site} \ll 1$.

\begin{figure}[h]
\centering    
\includegraphics[width=0.8\columnwidth]{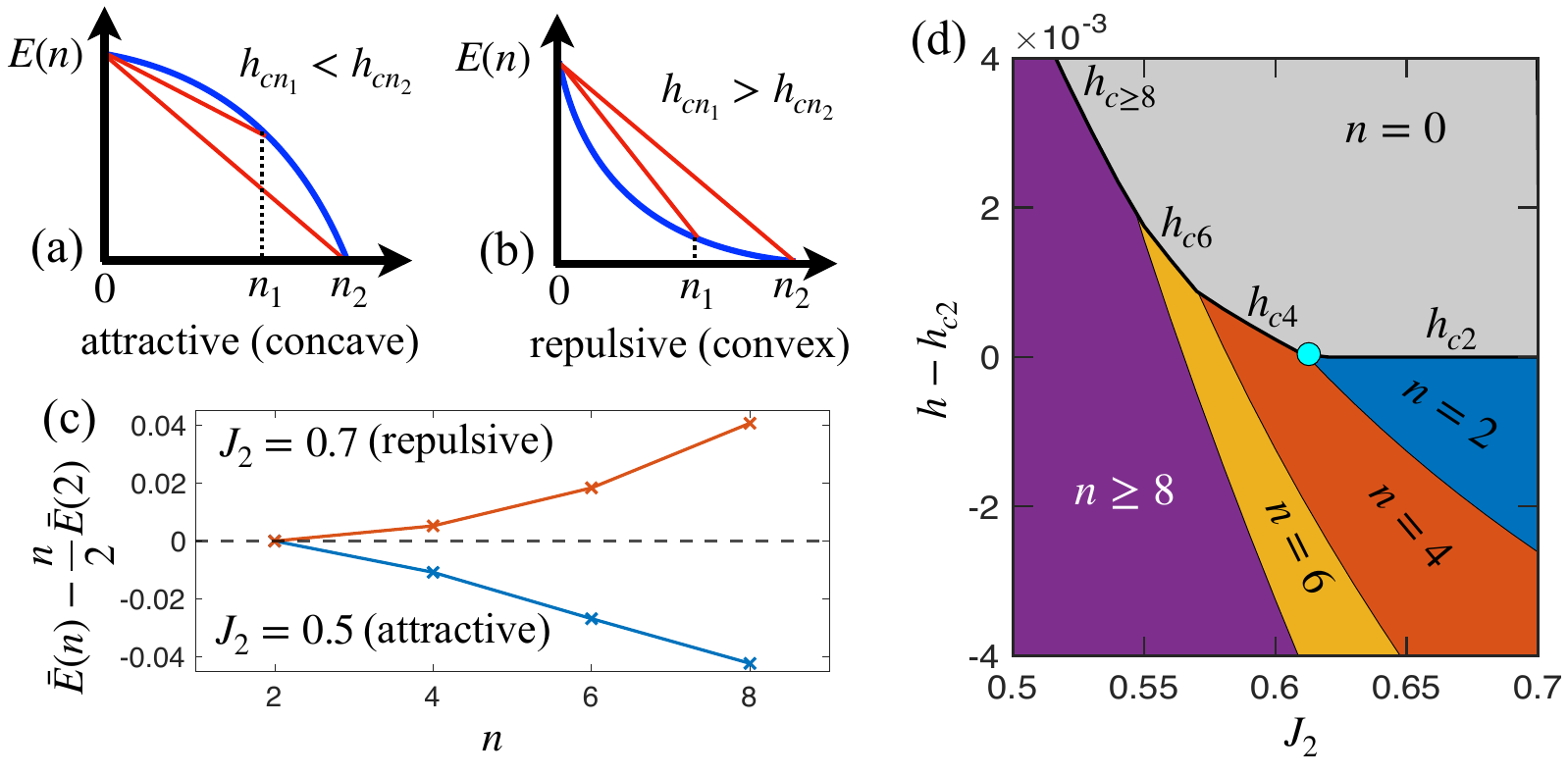}
\caption{(a) and (b) Schematic plots of the total energy $E(n)$ vs magnon number $n$ without the magnetic field.
The slope of the line connecting $E(0)$ and $E(n)$ is the multi-pair BEC instability field $h_{cn}$. The energy vs magnon number curve is concave (convex) for attractive (repulsive) interaction.
(c) The total energy of the system with $n$ magnons $\bar{E}(n)$ relative to the energy of $\frac{n}{2}$ non-interacting pairs $\frac{n}{2}\bar{E}(2)$ for $J_2=0.7$ and $J_2=0.5$ in a $16\times 8$ cylinder.
(d) The $h$--$J_2 $ phase diagram for the states with fixed number of magnons $n=2,4,6,8$ in a $16\times 8$ cylinder.  For $J_2>0.61$, magnon pairs are repulsive and the leading instability is the single-pair-BEC ($n=2$), while for $J_2<0.61$ the magnon pairs are attractive, which leads to the multi-pair instabilities with narrowing $J_2$ steps that resemble a devil's staircase.}
\label{fig:multipair}
\end{figure}

In Fig.~\ref{fig:multipair}(a) and Fig.~\ref{fig:multipair}(b), we show two schematic plots of the total energy of the system $E(n)$ vs magnon number $n$  for attractive and repulsive magnon pairs, respectively. The $n\!=\!0$ state is a fully polarized state and we do not include external field into consideration yet. 
In either case, the energy is lowered  by the increase of the magnon density because the fully polarized state is a highly excited state in the regime where an antiferromagnetic state is a ground state of the system in zero field. If  magnon pairs are attractive, the (negative) energy gain  by $n+2$ magnons is always greater than the sum of such energy gains by $n$ and $2$ magnons separately. Therefore, the $E(n)$ vs $n$ curve should be concave, as is shown in Fig.~\ref{fig:multipair}(a). In this case, the system can not be stabilized at a low magnon-pair density. Depending on the magnetic field, the ground state is either a saturated FM state or a multi-pair state with a large concentration of magnons, which is beyond the considered low-density regime. On the other hand, if the magnon pairs are repulsive, the energy curve should be convex, as is shown in Fig.~\ref{fig:multipair}(b), a state with a fixed magnon-pair density can be stabilized  by a field,  and the magnon pair-density in the ground state will increase continuously upon lowering of the field. In the following consideration, it is convenient to count the energy $E(n)$ from the energy $E(0)$ of the fully polarized state, $\bar{E}(n)\equiv E(n)-E(0)$.

In Fig.~\ref{fig:multipair}(c) we show the  energy of the system with $n$ magnons, $\bar{E}(n)$, relative to the energy of $\frac{n}{2}$ non-interacting pairs, $\frac{n}{2}\bar{E}(2)$, for $J_2=0.7$ and $J_2=0.5$,  obtained from DMRG calculations in a $16\times 8$ cylinder. For $J_2=0.7$, this energy is positive, indicating that adding another pair to the already existing ones cost extra energy due to repulsion between pairs. On the other hand, for $J_2=0.5$, this  energy is negative due to attraction between pairs.

To identify which multi-pair state has the leading instability with respect to the fully polarized state, we calculate and compare the $n$-magnon instability field $h_{cn}$, defined as $h_{cn}\!=\!\big(E(0)-E(n)\big)/n\!=\!-\bar{E}(n)/n$, for $n=2,4,6,8$ in a $16\times 8$ cylinder.
In the schematic plots in Figs.~\ref{fig:multipair}(a) and \ref{fig:multipair}(b), the values of $h_{cn}$ corresponds to the discrete versions of the (negative of the) slopes of the lines connecting $E(0)$ and $E(n)$. In the case of attractive magnon pairs, if the multi-pair states are formed that still repel each other, the instability field is achieved at $h_{cn}>h_{c2}$. The biggest achievable $h_{cn}$  corresponds to the leading instability.

In Fig.~\ref{fig:multipair}(d), we present the $h$--$J_2$ phase diagram showing such instability fields for the states with fixed number of magnons in a $16 \times 8$ cylinder  for $J_2$ from 0.5 to 0.7. To obtain this phase diagram, we calculated the total energy of the system in magnetic field $h$ as a function of $J_2$, $\widetilde{E}_n(h,J_2)=E_n(J_2)+nh$, where $E_n(J_2)$ is the total energy of the system with $n$ magnons without magnetic field as above. For each $(h,J_2)$ point in this diagram, the $n$-magnon state with the lowest $\widetilde{E}_n(h,J_2)$ is chosen. The phase boundary between the $n$-magnon state and the fully polarized state with $n=0$ corresponds to the $h_{cn}(J_2)$ discussed above.

For $J_2\gtrsim 0.61$, magnon pairs are repulsive and the leading instability field is $h_{c2}$. The states with more pairs are stabilized at a lower field.
For $J_2\lesssim 0.61$, the leading instability changes from $h_{c2}$ to $h_{c2n}$ in  steps of 2, with $h_{c2n}$ increasing as $J_2$ is decreased. At the same time, the $J_2$-ranges in which each $h_{c2n}$ dominates shrink with the increase of $n$,  resembling a devil's staircase of an infinite sequence of   transitions with the ever decreasing width. Although the shift of the leading instability away from the $h_{c2}$ line in favor of the multi-pair $h_{c2n}$ in our analysis is beyond doubt, the finite-size effects prevent a definite confirmation  of the devil's staircase scenario.

In the attractive regime there are two scenarios. If multi-pair objects repel, the system will be stabilized at $h_{cn}$. Near the boundary of the two-magnon stability ($J_2 \lesssim 0.61$), this enables nematic state to survive for a certain field range below $h_{cn}$ before entering the AFM state at a lower field. For a yet smaller $J_2$ ($J_2\alt 0.5$), and stronger attraction, the multi-pair state cannot be stabilized for any finite $n$ and the system undergoes  a first-order transition to a strongly canted AFM state. 
This transition is discussed in Sec.~\ref{j20.45} in more detail.

\section{reduction of the single-magnon gap in the presence of pairs}
\label{contraction}

As is discussed in the main text, the actual nematic phase  occupies only a fraction of the na\"{i}vely anticipated region between $h_{c1}$ and $h_{c2}$, with the rest of that region taken by the AFM phase. Since $h_{c1}$ corresponds to a closing of the single-magnon gap for the non-interacting magnons, an expansion of the AFM phase above $h_{c1}$ suggests that the single magnon-excitations can reduce their gap due to interaction with the existing pair-condensate.
Here, we substantiate this picture by providing  quantitative data for the reduction of  the single-magnon gap based on the DMRG calculations in the $16\!\times\!8$ cluster with fixed total $S^z$ for the representative value of  $J_2=0.7$.

\begin{figure}[h]
\centering    
\includegraphics[width=0.9\columnwidth]{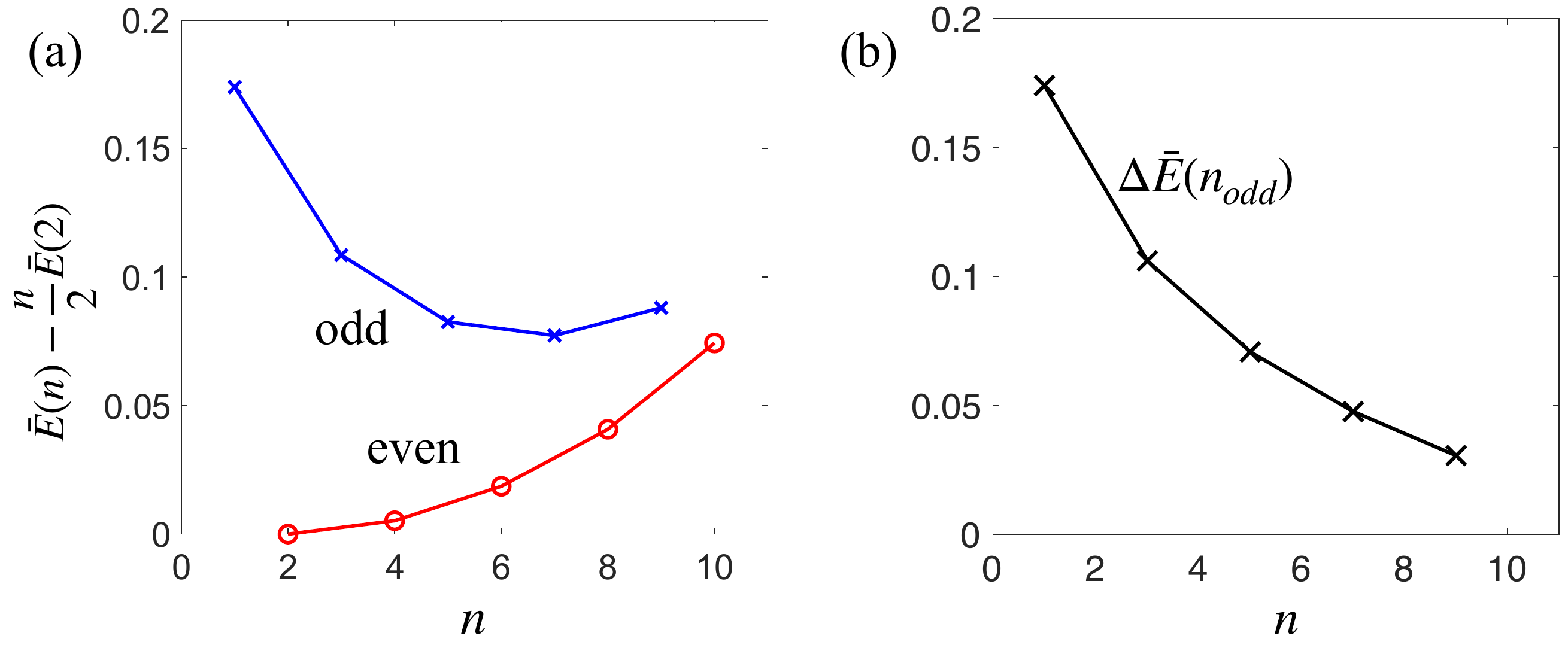}
\vspace{-0.3cm}
\caption{(a) The total energy of the system with $n$ magnons relative to the magnon pairing energy in a pair per magnon, $\frac12\bar{E}_2$, times $n$ vs $n$ in the $16\!\times\!8$ cluster with fixed total $S^z$ for $J_2=0.7$ and $n$ from 1 to 10. Red line is the even-$n$ sector and blue line is the odd-$n$ sector. (b) The single-magnon gap $\Delta E(n_{odd})$ vs $n$.}
\label{fig:EvsNfixedSz}
\end{figure}

First, we look at the total energy relative to the magnon pairing energy, as is shown in Fig.~\ref{fig:EvsNfixedSz}(a).
The total energy of the $n$-magnon state $\Bar{E}(n)$ is  the energy relative to the energy of the fully polarized state: $\Bar{E}(n)=E(n)-E(0)$. Considering it relative to the magnon pairing energy in a single pair per magnon times $n$, $\frac{n}{2}\Bar{E}(2)$, is similar to introducing a chemical potential, so that energies of the states with a different magnon number can be compared.
One can see in Fig.~\ref{fig:EvsNfixedSz}(a), that for the states with even values of $n$, in which  all magnons form pairs, the relative energy increases with $n$. This effect is simply due to repulsion between magnon pairs in a finite cluster. However, for the states with the odd $n$,  in which there is one unpaired magnon, their energy  decreases with $n$ despite the pair-repulsion, indicating an attraction of the single magnon to the pairs and a reduction of the single-magnon gap. 

In Fig.~\ref{fig:EvsNfixedSz}(b), we plot the single-magnon excitation gap, $\Delta \bar{E}(n)$, defined as $\Delta \Bar{E}(n)\!=\!(2\Bar{E}(n)-\Bar{E}(n-1)-\Bar{E}(n+1))/2$ for $n$=odd, which also corresponds to the average vertical distance between a blue cross and two closest red circles in Fig.~\ref{fig:EvsNfixedSz}(a).
Indeed, for $n\!=\!1$, the value $\Delta \Bar{E}(1)$ is nothing but the magnon pairing gap $\Delta$, see Fig.~2(a) of the main text, which separates the lowest magnon energy from that of the pair. Generalizing it to $n\!>\!1$, corresponds to the energy of an extra magnon with respect to the background energy of magnon pairs. As one can see in Fig.~\ref{fig:EvsNfixedSz}(b), this single-magnon excitation gap reduces significantly upon increase of $n$, implying that the magnon is attracted to the magnon pairs already at the level of one magnon--one pair, $n\!=\!3$, and lowers its energy  further by interacting with multiple magnon pairs for larger $n$.
The decrease of the single-magnon gap due to interaction with the condensate of magnon pairs results in an expansion of the AFM phase above the non-interacting phase boundary $h_{c1}$.

\section{finite-size scaling of the nematic order parameter}

To investigate  the order-parameter evolution with the system size, we follow the strategy of Ref.~\cite{FS}. It suggests that the DMRG cylinders with the open-periodic boundary conditions and the aspect ratio $L_x/L_y$ about 2.0 should cancel the leading finite-size effects for the order parameter in the center of the cluster, closely approximating  its value in the thermodynamic 2D limit. 

\begin{figure}[h]
\centering    
\includegraphics[width=0.7\columnwidth]{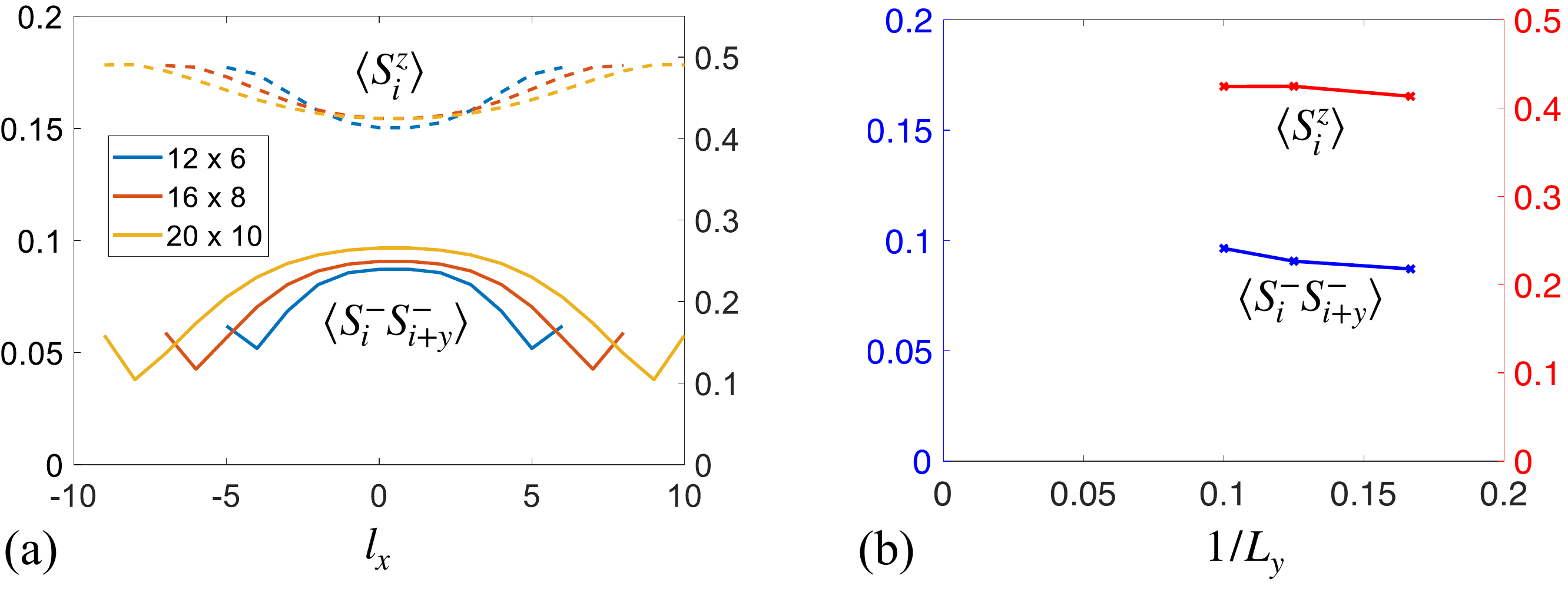}
\vspace{-0.3cm}
\caption{Finite-size scaling of the nematic order parameter and $z$-axis magnetization in the nematic phase for $h=0.96$ and  $J_2=0.7$. (a) Nematic order parameter $\langle S^-_i S^-_{i+y}\rangle$ (left axis) and $z$-axis magnetization $\langle S^z_i \rangle$ (right axis) along the cylinder for the three cluster sizes of the same aspect ratio 2.0. There is a weak symmetry-breaking pairfield $0.1 S^-_i S^-_{i+y}$ applied to both edges. (b) $1/L_y$-scaling of $\langle S^-_i S^-_{i+y}\rangle$ and $\langle S^z_i \rangle$ at the center of the cylinders in (a).}
\label{fig:scaling}
\end{figure}

 We use three cylinders with the aspect ratio of 2.0 and apply the nematic pair-field of 0.1$\langle S^-_i S^-_{i+y} \rangle$ on both edges. The results for the nematic order-parameter $\langle S^-_i S^-_{i+y}\rangle$ and for the $z$-axis magnetization $\langle S^z_i \rangle$  are shown in Fig.~\ref{fig:scaling}(a) as the profiles along the width of the clusters. The $1/L_y$-scaling of the nematic order parameter and magnetization in the center of the cylinders are presented in Fig.~\ref{fig:scaling}(b). Both quantities show little change, giving a strong support to the existence of the nematic phase in 2D and providing a vindication to the strategy of Ref.~\cite{FS} used in this work, suggesting that the width-8 cylinders that we rely on in our study approximate the 2D behavior very well.
 
\section{dipole and quadrupole order parameters in a  Heisenberg antiferromagnet}
\label{j1scan}

In a multipole expansion, higher-order multipole correlations can exist merely as a byproduct of the lower-order ones, so the nematic (quadrupolar) order parameter should always be detectable in the presence of the conventional dipolar order. In this section, we present a numerical demonstration of this effect  for the $J_1\!>\!0$, nearest-neighbor Heisenberg antiferromagnet in a magnetic field. By highlighting its differences from the $J_1$--$J_2$ ferro-antiferromagnetic model that we study in  the present work, we further confirm that the existence of the pure nematic phase in the $J_1$--$J_2$ model is {\it not} a byproduct of the dipolar order.

\begin{figure}[h]
\centering    
\includegraphics[width=0.7\columnwidth]{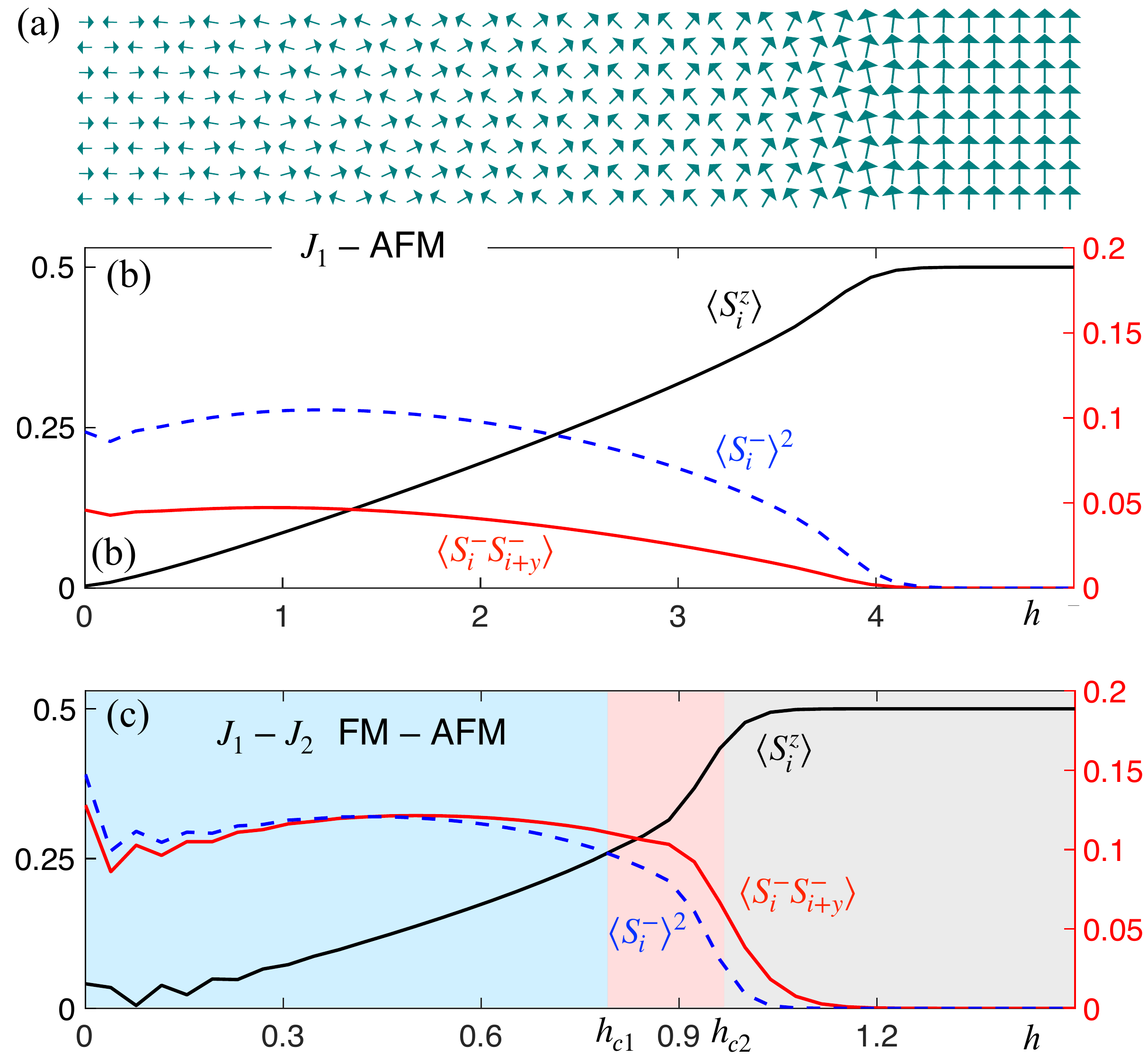}
\caption{(a) Spin configuration in a long-cylinder scan of the Heisenberg $J_1$-AFM, with the magnetic field varying linearly across the length of the cylinder from zero to above the saturation field. (b) Magnetization $\langle S^z_i \rangle$ (left axis),  square of the dipolar order parameter, $\langle S^-_i \rangle^2$,  and the quadrupolar order parameter $\langle S^-_i S^-_{i+y} \rangle$,  (right axis) along the cylinder in (a). (c) The same order parameters as in (b) for the $J_1$--$J_2$ FM-AFM model for $J_2=0.7$, as in Fig.~4(b) of the main text, but for a wider field range. The shaded regions are the phases from the naive phase diagram in Fig.~1(a) of the main text based on the values of $h_{c1}$ and $h_{c2}$. The actual AFM to nematic phase boundary is at $h \approx 0.95$, which is way above $h_{c1}$ and only slightly below $h_{c2}$.}
\label{fig:j1scan}
\end{figure}

In Figs.~\ref{fig:j1scan}(a) and \ref{fig:j1scan}(b) we show the DMRG results for the long-cylinder scan  of  the Heisenberg $J_1$-AFM in a magnetic field, with $J_1=1$ set as the energy unit. The magnetization curve shown in Fig.~\ref{fig:j1scan}(b) agrees very closely  with the prior spin-wave calculation \cite{j1field}.
One can see that the spin canting, demonstrated by the square of the order parameter, $\langle S^-_i \rangle^2$, occurs  below saturation field with no sign of an intermediate nematic phase. There is also a finite quadrupolar order parameter, $\langle S^-_iS^-_{i+x(y)} \rangle$, which is detected alongside the dipolar order, but not independently.  Furthermore, the dipolar order is always stronger than the quadrupolar one throughout our scan, with no region where the former can dominate and indicate a possible occurrence of the pure nematic state. 

This should be contrasted with the scan of the $J_1$--$J_2$ ferro-antiferromagnetic model for  $J_2=0.7$, as in Fig.~4(b) of the main text, but for a wider field range, shown in Fig.~\ref{fig:j1scan}(c). One can observe a region near the saturation field where the ordered moment has deviated from the full saturation, but shows no spin canting, also indicated by the dominant quadrupolar order, $\langle S^-_iS^-_{i+y} \rangle$, with the suppressed or vanishing dipolar one, $\langle S^-_i \rangle^2$, in this region. These are the distinct features of the spin nematic phase as we argue and analyze in more detail in the main text. 

\section{first order transition from FM to AFM at $J_2=0.45$}
\label{j20.45}

\begin{figure}[t]
\centering    
\includegraphics[width=0.7\columnwidth]{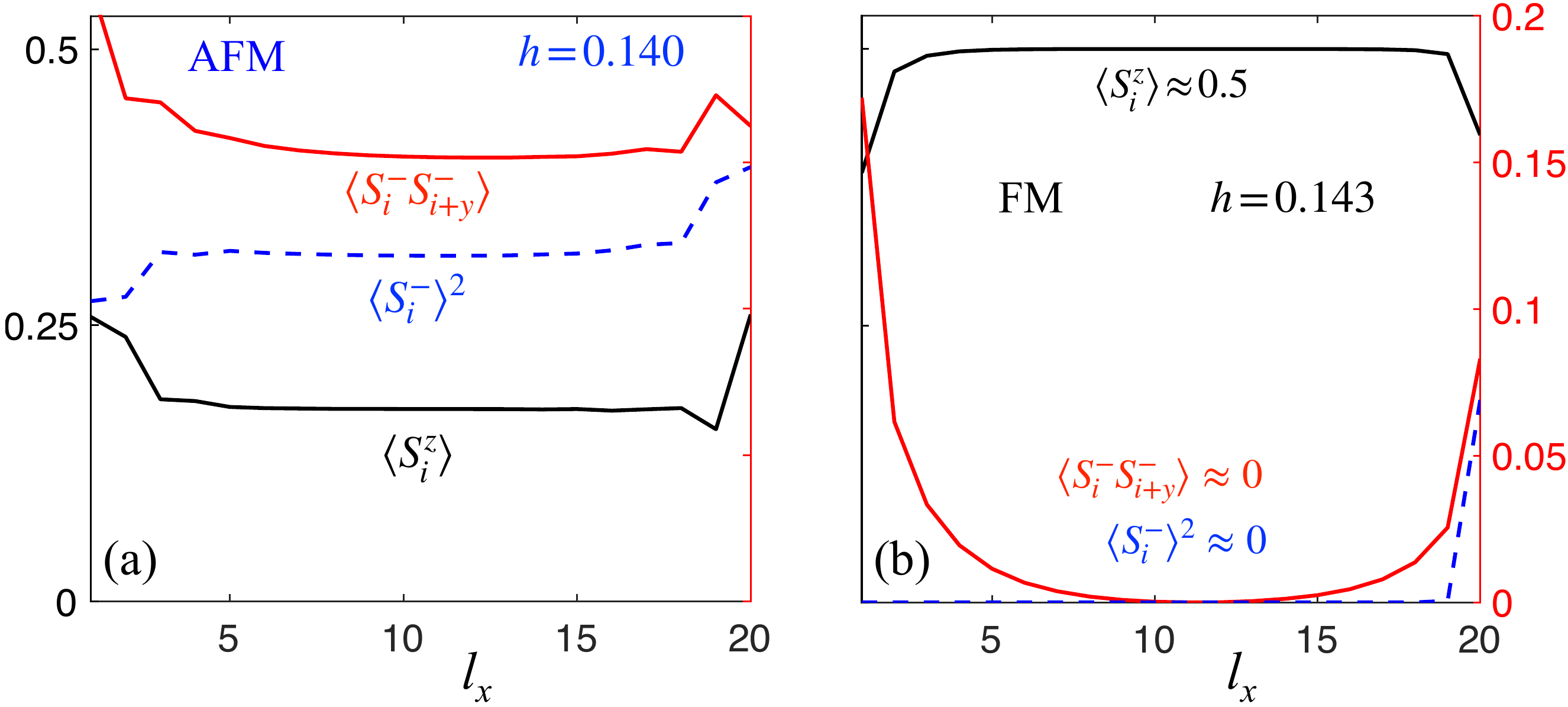}
\caption{DMRG non-scan results for $J_2=0.45$, see Fig.~5 in the main text. (a) The ground state at $h=0.140$ is a strongly canted AFM state as is indicated by the magnetization $\langle S^z_i \rangle$ (left axis), and by the dipolar, $\langle S^-_i \rangle^2$,  and quadrupolar, $\langle S^-_i S^-_{i+y} \rangle$, order parameters (right axis). (b) The $h=0.143$ state is a fully saturated FM state. We apply small pairing and canting fields on the left and right edges, as is described in the main text.}
\label{fig:j20.45}
\end{figure}

In this section, we provide further evidence supporting the first-order nature of the FM to AFM transition at $J_2$=0.45, based on the calculations in the  clusters with fixed parameters (i.e., non-scan calculations).
As is shown in Fig.~\ref{fig:j20.45}, for the two very close magnetic fields, $h=0.140$ and $h=0.143$, the ground states are very different.  For $h=0.140$, it is a strongly canted AFM state, while for $h=0.143$ it is a fully saturated FM. This discontinuity in the ground state spin configuration is a clear indication of the first-order transition,  consistent with the scan calculations, shown in Fig.~5 of the main text. 

Another general feature of the first-order transition is a possibility of the metastable states, in which the two states that are separated by such a transition can appear stable beyond the regions where they constitute global energy minima. We can  observe such a metastability in DMRG calculations. As was mentioned in Sec.~\ref{multpair}, in the  attractive-pair regime, the ground state is either a saturated FM state or a strongly canted AFM state with a large magnon density. An implication of this two drastically different outcomes is that these states remain  local minima in terms of magnon density, even if they are not true ground states.  For example, in the $h=0.143$ case (polarized FM ground state), we can initiate DMRG calculation in the collinear N\'eel state, and the cluster will be stuck in the metastable canted AFM state, showing correlations that are similar to the one in Fig.~\ref{fig:j20.45}(a), but having a higher energy.  On the other hand, for the repulsive-pair regime ($J_2>0.6$), where all phase transitions are expected to be continuous, we do not detect such a metastability in our DMRG calculations, in agreement with these expectations. 


\bibliography{ref}